# Phase Transition of Iron-based Single Crystals under Ramp Compressions with Extreme Strain Rates

Kun Wang[a], Jun Chen[a,b*], Wenjun Zhu[c*], Wangyu Hu[d], Meizhen Xiang[a]

[a] *Laboratory of Computational Physics, Institute of Applied Physics and Computational Mathematics, Beijing 100088, PR China*
[b] *Center for Applied Physics and Technology, Peking University, Beijing 100071, China*
[c] *National Key Laboratory of Shock Wave and Detonation Physics, Institute of Fluid Physics, Mianyang 621900, China*
[d] *College of Materials Science and Engineering, Hunan University, Changsha 410082, China*

## Abstract

Recent widespread interests on strain rate effects of phase transition under dynamic loadings bring out present studies. α→ε phase transition of iron, as a prototype of martensite phase transition under dynamic loadings, exhibits huge diverges in its transition pressure (TP) among experiments with different pressure medium and loading rates, even in the same initial samples. Great achievements are made in understanding strain or stress dependence of the TP under dynamic loadings. However, present understandings on the strain rate dependence of the TP are far from clear, even a virgin for extreme high strain rates. In this work, large scale nonequilibrium molecular dynamics simulations are conducted to study the effects of strain rates on the phase transition of iron-based single crystals. Our results show that the phase transition is preceded by lattice instabilities under ramp compressions, but present theory, represented by modified Born criteria, cannot correctly predict observed onsets of the instability. Through considering both strain and strain gradient disturbances, new instability criteria are proposed, which could be generally applied for studying instabilities under either static or dynamic loadings. For the ramp with a strain rate smaller than about $10^{10}$ s$^{-1}$, the observed onset of instabilities is indeed equal to the one predicted by the new instability criteria under small gradient disturbances. The observed onsets deviates from the predicted one at lager strain rates because of finite strain gradient effect —— nonzero higher order stresses and work conjugates of the strain gradient. Interestingly, the strain rate ($\dot{\varepsilon}$) dependence of the TP also exhibits an obvious change at the same strain rate, i.e., $10^{10}$ s$^{-1}$. When $\dot{\varepsilon} \leq 10^{10}$ s$^{-1}$, a certain power law is obeyed, but it is not applicable at larger strain rates. This strain rate effect on the TP are well interpreted with nucleation time and the finite strain gradient effect. According to these basic understandings, the roles of strain rates on nucleation and growth of the phase transition are studied. Besides, initial shock formation time at extreme strain rates is analyzed, which also exhibits a deviation from a scaling law.

**Key Words:** Phase transition; Strain rate; Strain gradient; Lattice instability; Molecular dynamics; Ramp compressions; Metal.

---

# 1. Introduction

Phase transition under high pressures has been attracting a great interest in condensed matter physics (Duvall and Graham, 1977; Kadau et al., 2002; Takahashi and Bassett, 1964), geophysics(Coppari et al., 2013; Shen et al., 2016), materials science(Li et al., 2014; Talonen and Hänninen, 2007) and engineering mechanics(Guthikonda and Elliott, 2013). Recent breakthrough on the ultrafast X-ray diagnostics (Denoeud et al., 2016; Milathianaki et al., 2013) under dynamic loadings has prompted a renewed interest in phase transformation of iron from α (bcc) to ε (hcp) phase which is a prototype of martensite phase transition under high pressures. In the past thirty years, mechanical behaviors of materials exhibiting martensite phase transition have been simulated with methods at different hierarchical levels from macroscopic phenomenological models to microscopic-mechanism-based models. As reviewed by (Fischlschweiger et al., 2012), the majority of the models use the martensite fraction as a key variable whose time evolution rules are known as the phase transition kinetics. Typical phase transition kinetics could be either controlled by stress or strain according to the physical nature of the phase transitions. Systematic studies on these two types of phase transitions have been done by (Levitas, 2004a; Levitas, 2004b; Levitas, 2004c). Except for the phase transition kinetics, numerically modeling phase transition under external loadings also requires a transition criterion as well as critical forces of the phase transition, for example a recent developed semi-phenomenological model by (Kubler et al., 2011). The transition criterion could be built up via transition driving force from parent phase to product phase if local thermodynamic equilibrium states could be set up. However, under extreme strain rates, this transition criterion may be not valid because of the extreme non-equilibrium nature, and can no longer be studies on a macroscopic level. The behaviors of materials exhibiting phase transition in response to the extreme strain rates, for example, shock compressions, could be investigated through large scale molecular dynamics simulations, which have shown to be highly intricate even in defect-free systems (Branicio et al., 2013). In this work, atomic-simulation-based analyses on the phase transition are utilized to study the transition criterion as well as the critical forces or transition pressure (TP).

Since martensite phase transition of metals under shock compressions is discovered in iron for the first time (Bancroft et al., 1956; Takahashi and Bassett, 1964), quantities of experiments are conducted under both dynamic and static compressions to uncover its existence and study its dynamics. It is believed that bcc↔hcp phase transition of iron under dynamic loadings happens at a metastable two-phase surface where actual mass fraction of ε phase need a time, termed relaxation time, to develop to the one of thermodynamic equilibrium (Boettger and Wallace, 1997). An outstanding character of the metastable state is "over-pressurization" beyond the equilibrium phase boundary at high strain rate. It is found that the relaxation time could relate to the over-pressurization approximately through an exponential function (Jensen et al., 2009). Hence, the knowledge of the over-pressurization, or alternatively the onset of the phase transition, is crucial to understanding the dynamics of the phase transition. Historically, the transition pressure of iron detected in different experiments varies from about 8 GPa to 25 GPa (Amadou et al., 2016; Crowhurst et al., 2014; Zarkevich and Johnson, 2015), which has been attributed to reasons of initial shear of samples, pressure mediums and local strain or stress states. Besides, the TP exhibits a strong strain-rate ($\dot{\varepsilon}$) dependence under extreme strain rate (Amadou et al., 2016; Crowhurst et al., 2014; Jensen et al., 2009; Smith et al., 2013). Smith et., al. (Smith et al., 2011) found that the rate-dependence TP of iron is similar to the rate-dependence plastic flow observed in aluminum and iron governed by the dislocation flow mechanism, which satisfies a power law when $\dot{\varepsilon}>10^6$ s$^{-1}$ (Smith et al., 2013). Other studies (Pang et al., 2014a, b; Wang et al., 2014; Wang et al., 2015) suggests the time, needed to form a nucleus, may be responsible to the over pressurization under high strain rate. Recent analysis of strains and stresses on fast compressed iron shows that the c/a ratio of hcp iron changes with the strain rates, as well as relation time and various levels of hydrostaticity (Konôpková et al., 2015), which indicates a more complex physical picture of the dynamic phase transition than the Bain-path based understandings. Despite of these recent awareness on the strain-rate dependence of the TP, the detailed physical picture of rate-dependence phase transition at lattice level is still unclear, for example, what roles the strain rate plays in dynamic phase transition and how to understand the loading rate dependence of c/a ratio of hcp iron from the phase transition mechanism found in the shock compressions (Hawreliak et al., 2006; Kadau et al., 2005; Kadau et al., 2002; Wang et al., 2014). Previous molecular dynamics studies on the shock induced phase transition of iron suggest that lattice instability of the over-pressurized bcc iron may be responsible to the nucleation of hcp phase (Shao et al., 2009; Wang et al., 2015). The lattice instability under complex stress or strain states could be predicted by Born's criteria or the modified B criteria (Wang et al., 1995; Wang et al., 1993), which is corresponding to upper limit of TP (Morris Jr et al., 2001). In facts, the Born's criteria is a special case of a more general criteria —— phonon instability criteria in the long wave length limit of the phonon dispersion curves, which has been systematically reviewed in reference (Grimvall et al., 2012). In addition, the contribution of vibrational entropy to stiffness coefficients may play a critical role in determining the instability limit of crystals under the high-temperature high-pressure conditions (Kong et al., 2012). It is worth noting that neither of these methods is rate-dependence which result in the essential difficulties in understanding the rate dependence of the phase transition under dynamic loadings. Besides, the nucleation of hcp phase will begin after bcc iron becomes instable, but dynamic pictures on how nucleation proceeds in the instability region are still lack. Since the shear (or compression) and shuffle part of the phase transition mechanism of bcc iron are essential processes for the nucleation of hcp phase, some valuable insight on the nucleation mechanism could be obtained by studying the detailed transition path from bcc to hcp phase via different sequences of the two processes. A*b* initio calculations (Dupé et al., 2013; Lane et al., 2016; Lu et al., 2014) show that the detailed transition path relies on pressure of initial bcc iron, and, under a certain fixed shear mode, a consecutive shuffle manner is more energetically favorable than simultaneous shuffle. However, the nucleation begins in a much more complex environments under dynamic loading, which may be not a simple coupling mode between the two processes of the phase transition of iron as suggested in recent studies on the shock induced phase transition of polycrystalline iron (Wang et al., 2015). Nucleation and growth of the hcp phase under uniaxial uniform compressions upon single crystalline iron are widely investigated through molecular dynamics simulations (Pang et al., 2014a, b; Shao et al., 2009; Wang and et al., 2009, 2010). It is generally recognized that the phase transition begins with homogeneous nucleation and exhibits a supersonic growth speed at the early stage of the nucleation. Despite of some similarities shared between the uniaxial uniform compressions and uniaxial shock compressions, there are still some distinctive features for the phase transition under dynamic loadings due to the dynamic nature, characterized by wave propagations, for example, variant selections of the transition product (Kalantar et al., 2005; Kanel et al., 2015; Wang et al.,

2014; Wang et al., 2015).

In contrast to traditional dynamic (shock) compressions, ramp compression is a kind of quasi-isotropic (QI) compression technique (Amadou et al., 2013; Martin et al., 2012; Seagle et al., 2013), which could generate a pressure range up to the order of terapascal in solid matter (Smith et al., 2014). Because pressure range generated by the ramp compression is much larger than that produced by static high pressure techniques (such as diamond anvil cell), it is broadly employed in the matter research regime under extreme conditions (Coppari et al., 2013; Smith et al., 2014; Xue et al., 2016). Combined with the shock compressions (a so-called shock-ramp compressions), different thermodynamic compression paths of materials, including the P-T region between the shock Hugoniot and isentrope, are allowed to be accessed and thus open a much wider thermodynamic space for materialists to explore (Jue et al., 2013; Seagle et al., 2013). Besides, the ramp compression is adopted for purpose of effective performances in inertially confined fusion implosions, either direct (ablative) or indirect (hohlraum) drive (Swift and Johnson, 2005). Unfortunately, the ramp compression wave in materials is unstable, which will progressively steepen and finally forms a shock after a certain propagation distance. This means that only materials within the distance are under isotropic compressions. According to theoretical model of Swift and et. al. (Swift et al., 2008), shock-formation time, closely related to the propagation distance, is an order of 1.5 times of ramp rising time over a wide range of strain rates. Higginbotham and et. al. (Higginbotham et al., 2012) find that the time predicted by the model is slightly smaller than the one for strain rates larger than $10^9$ s$^{-1}$ through atomic simulations. However, in their studies, solid-solid phase transition is not involved. It is found that the solid-solid phase transition favors the shock formations during ramp compressions and results in deviations from expected thermodynamic path (Amadou et al., 2013; Morard et al., 2010), but the formation processes of the shock at lattice level are not clear at present. Understanding the shock-formation processes during ramp compressions upon material exhibiting solid-solid phase transitions would prompt a better application of this technique in related research regime, which prompts one of interests of present studies.

In this work, nonequilibrium molecular dynamics (NEMD) simulations, combined with a recently developed interatomic potential of iron, are conducted to study the bcc↔hcp phase transition of iron under ramp compressions with different applied strain rates. Several interesting phenomena are observed in the simulations, for example, lattice instabilities take place before the phase transition and an initial shock forms before finial steady shock wave. However, the instabilities cannot be explained by the modified Born criteria developed for static compressions. This is to say, dynamic and static instabilities have different physical natures, although both of them are caused by deformations. Here, strain gradient disturbances due to the applied strain rates are believed to be a key factor that discriminate static compressions from the dynamic one. Then the instabilities are theoretically analyzed in terms of strain gradients, which show an amazing consistence between the simulation results and the theoretical one. The analyses on the instabilities indicate that the concept of the strain gradient is of crucial importance for understanding dynamic mechanical behaviors of metals at lattice level under extreme strain rates. With the concept, onset of the phase transition, nucleation before and after the formation of the initial shock and the initial-shock formation time at extreme strain rates are well interpreted. The results of this work could be expected to give some new insight into the physical nature of the phase transition under dynamic loadings, and, perhaps, to provide clues for interpreting continuum phenomenon from atom scale.

## 2. Theoretical Methods

NEMD simulations are conducted to study the phase transition of single crystalline iron under ramp compressions. The ramp compressions are performed by compressing an iron sample through a rigid wall which moves linearly from 0 km/s to $v_{max}$ (km/s) within a time of $t_{rising}$. After the ramp rising time ($t_{rising}$), the iron sample is compressed with a constant velocity of $v_{max}$, which will create a final particle velocity of $v_{max}$ in the sample. The ramp compression will create a strain rate ($\dot{\varepsilon}$) of about $\dot{v}_p/C = v_{max}/(Ct_{rising})$ before the sample yields, where $C$ is speed of elastic wave and $\dot{v}_p$ is time derivative of particle velocity $v_p$. Because $C$ is a material constant for a certain compression, the ratio of $v_{max}/t_{rising}$, or $C\dot{\varepsilon}$, is used to represent the applied strain rate in this work. It is noticed that this applied strain rate has the same dimension with acceleration, which is reasonable because one could control an applied strain rate by modifying accelerations of the moving wall. Modified analytic embedded-atom-model potential is employed to describe the interatomic interactions of iron, which is specifically developed for simulating the phase transition of iron under shock compressions (Wang et al., 2014). To observe the phase transition of iron, $v_{max}$ should be greater than or equal to 0.5 km/s for the interatomic potential of iron adopted here. Hence, we perform ramp compressions along [001] direction of perfect bcc iron at 0K, with a maximum particle velocity of 0.8 km/s and a ramp rising time of 5, 10, 15 and 50 ps, respectively. As a comparison, another group of ramp compressions, consisting of three simulations, is performed by fixing the ramp rising time at 50 ps while increasing the maximum particle velocity from 0.8 to 1.5 km/s (that are 1.0, 1.2 and 1.5 km/s). The dimension of initial iron sample is 14.30×14.30×286.06 nm$^3$ (total 5 000 000 atoms). A slightly enlarged iron sample, whose dimension is 17.16×17.16×286.06 nm (total 7 200 000 atoms), is employed for the second group of ramp compressions. Large-scale Atomic/Molecular Massively Parallel Simulator (LAMMPS) is used for the NEMD simulations, which enable us to track time evolutions of position, as well as velocity, of atoms in the simulated sample. In this work, all of the simulated systems run in the microcanonical ensemble with shrink boundary in Z direction and periodic boundary condition in the transverse directions. Motion equations of atoms are integrated by the velocity-Verlet algorithm with a time step of 0.4 fs and 0.5 fs for the first and the second group of simulations, respectively. Local lattice structure of our simulated samples is identified by the adaptive common neighbor analysis (Stukowski and Arsenlis, 2012) and coordination number. Wave profiles, characterized by certain local physical properties (such as density, particle velocity and the *ZZ* component of local stresses), are obtained by dividing the simulated sample of interest uniformly into many bins (with a width of about 8 Å for each bin) along Z direction and averaging the local properties within each bin. Analysis methods for specific problems, concerned in this work, will be discussed in the next part in details. Additional derivations of formulas used for our analyses are put in the Appendix A and B.

## 3. Results and Discussions

## 3.1 Phase Transition and Wave Propagations in Single Crystalline Iron

Wave profiles, represented by particle velocity ($v_p$) and ZZ component of local stresses, are calculated at several typical moments for different ramp compressions (See Fig. 1 and S1-3). As shown in Fig. 1 a-b, structures of the wave profiles changes with wave propagation time before forming a steady shock wave during ramp compressions. To understand the microscopic mechanism behind the changes of the wave structures, spatial distributions of the first nearest neighbor separation distance ($r_0$) and coordination number along Z direction are analyzed at the moments corresponding to that of the particle velocity profiles (See Fig. 1 c-d). Our cutoff distance, used for calculating the coordination number, is set to be 2.7 Å, which locates between the first and the second nearest neighbor separation distance of bcc iron (as well as that of the phase transition products). According to comparisons of this results with that obtained by the adaptive common neighbor analysis method, the coordination number is 8 for bcc atom, and 12 for fcc and hcp atom. For the shock along [001] direction of bcc iron, fcc atoms just appear as stacking faults or grain boundary atoms (Wang et al., 2014), whose amounts could be neglected when compared with that of the hcp atoms. In other words, atoms with a coordination number of 12 could be considered to be hcp atoms in our results. Other values are corresponding to some structures which may be not well-defined and, in most case here, are lattice defects, such as grain boundaries and the embryo nucleus of the transition products (See Fig. 2). From the results shown Fig. 1-2, several features could be identified. First, homogeneous nucleation begins after bcc phase changes into a ten-coordination-number (TCN) structure during ramp compressions. That is to say, the hcp phase is transformed from the TCN structure rather than a bcc phase. The formation of the TCN structure is due to small relaxations of instable bcc phase via compressions along [001] direction within (110) or (1-10) plane and forming a hexagon pattern in these planes. The TCN structure is also observed before the phase transition in the shock upon single crystalline iron (Wang et al., 2014) where the TCN structure is corresponding to the result of the first step toward phase transition of iron from bcc to hcp phase. The emergence of the TCN structure is due to lattice instabilities of the compressed bcc phase, which will be discussed in details in the next section. As shown in Fig. 1, the change from compressed bcc phase to the TCN structure does not cause an apparent rising in temperature, while the transformation from the TCN structure to hcp phase does. And the phase transition of iron makes the compression wave quickly become a shock. Second, the phase transition wave (TW), represented by the moving phase interface, proceeds in a region consisting of the TCN structure. For brevity, the region consisting of the TCN structure is called instability region and the boundary between the instability region and the bcc phase region is called instability boundary (IB) in present work. From snapshots of compressed iron sample shown in Fig. 2, nucleuses of hcp phase would be continuingly created in the instability region and later, be combined by the passing-by phase interface. The right boundary of the nucleation region, or NB for short, propagates faster than the IB at early stages of ramp compressions, which would eventually catch up to the IB after a certain wave propagation time and thus form an initial shock. If the finial particle velocity is large enough (as the case of present work), the TW would catch up to the NB after the initial shock forms, which results in a steady single shock wave. Formation time of the initial shock grows with the decreasing of the applied strain rates, which will be discussed further in section 3.4. Among the ramp compressions performed here, when the applied strain rate is smaller than 0.8 Å/ps$^2$, the final steady shock wave cannot be observed within the allowed wave propagation time due to the limited material dimension along Z direction. Otherwise, the finial steady shock wave could be observed. Fig. 3 has shown the formation processes of the finial steady shock wave for an applied strain rate of 1.6 Å/ps$^2$. The movement of the phase interface is driven by the growth of the transition product and the combinations with the nucleuses generated in the instability region. While the movement of the NB is driven via interactions between nucleation near the NB and the strain fields generated by nearby nucleus, which will be addressed further in section 3.3.

It is known that lattice instability is determined by the strain states of the sample and applied strain rate is equivalent to an additional strain gradient applied to the compressed materials under dynamic loadings. To understand strain rate effects of the instability, it is instructive to study spatial distributions of strain and its gradient in the sample under different applied strain rates. Compressed samples, analyzed in Fig. 4, are corresponding to the moments when the samples begin to be instable at somewhere near the compressed end. Because the strain cannot be well-defined at lattice level after the occurrence of the instability, the spatial distribution of the strain only ranges from the position, where the instability begins, to the free end of the compressed sample. If not specified, finite Lagrangian strain is employed for the strain analyses. The distributions of corresponding strain gradients for different ramp compressions are also shown in Fig. 4. These results indicate that a constant applied strain rate does not necessarily generate an ideal constant strain gradient in the compressed sample before the instability happens because of the nonlinear elastic response to the ramp compressions. General responses of the strain gradients to different constant applied strain rates could be states as follows. The constant applied strain rate will generate a constant strain gradient in the compressed bcc iron when the strain is smaller than about -0.017 (as marked by "A" in Fig. 4). The continuing compressions with the applied strain rates will bring the compressed bcc iron to a nonlinear-elastic-response region and thus leads to a rising in the strain gradient until the instability happens. Under the extreme applied strain rates, the generated strain gradient in the samples could not be neglected in analyses of the instability, as well as the phase transition kinetics. In the remaining part of present work, the strain gradient will be employed to interpret the above observed phenomenon.

*3.2 The Onset of the Phase Transition under Ramp Compressions*

Lattice instabilities due to the compressions along <001> direction are believed to be responsible to the α→ε phase transition of single iron crystal under shock compressions at low temperature (Kalantar et al., 2005; Wang et al., 2014; Wang et al., 2015). However, the detailed reason that triggers the instability under dynamic loadings is still unknown. In present work, we explore the reason of the instability happening under ramp compressions by comparing critical strain (at which the single iron crystal begins to become instable) obtained through two approaches. In the first approach, the critical strain is directly measured from a sample to be instable under ramp compressions. The strain at the position where the instability firstly begins is the critical strain. The measured critical strain versus the corresponding strain gradient for different ramp compressions is shown in Fig. 5a. In the second approach, the critical strain is evaluated by gradually compressing bcc-iron along [001] (Z) direction at 0 K and checking its stabilities via the modified B criteria (Wang et al., 1995; Wang et al., 1993). A more general way to obtain the modified B criteria is given in Appendix A. The strain instability is checked in terms of minimum eigenvalue of modified elastic stiffness matrix **B**, defined by (A.11'), under finite-strain configuration as a function of uniaxial strains. For brevity, the absolute value of strain, where $B_{min} = 0$, is termed critical strain. As shown in Fig. 5a and 5b, the critical strain is predicted to be 0.128 (marked by "B" in Fig. 5b) by the modified B criteria under uniaxial compressions. However, the critical strain (0.086) under ramp compressions is much smaller than the predicted value, and would be obviously decreases when the strain gradient is larger than 1.43e-4 Å$^{-1}$. It is known that the critical strain (or ideal yield limit), predicted by the modified B criteria, could be reached in perfect lattice at 0 K under quasi-/static compressions. According to our results, the ideal yield limit of perfect lattice under quasi-static compressions is not equal to the one under ramp (dynamic) compressions. To further demonstrate this point, we additionally investigate the critical strain of perfect iron single crystals under homogenous uniaxial compressions along [001] direction with a strain rate ($\dot{\varepsilon}$) ranging from $10^8$ to $10^{12}$ s$^{-1}$. The discriminate engineering strain from the finite strain, we have use ε to denote the former while η to denote the later. Dimension of the initial iron sample is 14.303 ×14.303 ×22.885 nm$^3$ (total 40 000 atoms). The simulations are performed in micro-canonical ensemble with a time step of Δ*t*, satisfying $\dot{\varepsilon}\Delta t = 10^{-6}$ for $\dot{\varepsilon} \geq 10^9$ s-1, and 2 fs for $\dot{\varepsilon} = 10^8$. Stress-strain curves at different strain rates are shown in Fig. 6, where the critical strain is nearly equal to the one predicted by the modified B criteria when $\dot{\varepsilon} \leq 10^9$ s$^{-1}$, while it monotonically increases with the growth of the strain rate. This growing trend is obviously different from that observed in the ramp compressions. A distinct difference between dynamic compressions and static compressions is whether waves are generated during the compressions. In present work, the ramp compression belongs to the dynamic one while the homogenous uniaxial compression belongs to the static one. These results again infer that the dynamic yield limit is different from the static one. According to the analyses above, the nonhydrostatic stress or uniaxial strain states can only explain the yield limit under the quasi-/static compressions although they are found to be important at other situations under extremely strain rates (Levitas and Ravelo, 2012; Zarkevich and Johnson, 2015). Under dynamic compressions, wave propagation will surely generate additional strain gradient in compressed samples, especially at the wave front. The strain gradient grows with the increment of applied strain rate. Thereby, the strain gradient, rather than the strain rate, is the key factor that determines the dynamic yield limit. Moreover, the strain gradient, like strain or volume, could be viewed as a state quantity, and thus conveniently relates to physical qualities of crystals via energy. Below, a general theory on strain-gradient instability (SGI) criteria will be developed through thermoelastic approaches. Since the SGI criteria are built at continuum level, its evaluation at atomic level is nontrivial. For such reason, an atomic-level approach to the SGI criteria is developed, which is verified to be consistent with the continuum one. With the atomic-level approach, the dynamic yield limit of single crystalline iron is predicted by the SGI criteria, and then make comparisons with the one obtained from NEMD simulations in order to verify the correctness of the theory. To make present work more compact, main theoretical derivations are put in *Appendix A* and *B*, while key points of the theory relevant to this work are discussed in this part.

    Supposing that a crystal is strained from an initial configuration {**a**} to a current configuration {**X**} via a strain and strain gradient, then we will discuss the stability of the crystal in {**X**} under a small virtual strain gradient which will take the strained crystal from {**X**} to configuration {**Y**}. Here, the applied strain and strain gradient are viewed as a disturbance. By convention, we use a black bold letter to denote a tensor (or a vector) and a tilt letter with

subscripts to represent the corresponding components. If not specified, summation over repeated indexes is adopted throughout this work, and $i, j, k, l, m, n$ ($I, J, K, L, M, N$) denote Cartesian indexes. The finite Lagrangian strains, measured with respect to $\{X\}$, are defined as

$$\eta_{ij} = \frac{1}{2}\left(X_{K,i}X_{K,j} - \delta_{ij}\right). \tag{1}$$

where $\delta_{ij}$ is the Kronecker delta. The deformation gradient tensor $X_{i,j}$ is related to $\{a_i\}$ and $\{X_i\}$ by

$$X_I = X_{I,k} a_k. \tag{2}$$

The second gradient of displacement $u_{k,ij}$ and the gradient of strain $\kappa_{ijk}$ could be defined as

$$u_{k,ij} = \partial_i \partial_j u_k = X_{k,ij}, \tag{3}$$

$$\kappa_{ijk} = \partial_i \eta_{jk} = \eta_{jk,i}, \tag{4}$$

where $u_k = X_k - a_k$, and $\eta_{jk,i}$ is the gradient of Lagrangian strain. The definition is Parallel to that of Mindlin and Fleck (Fleck and Hutchinson, 1997; Mindlin and Eshel, 1968) under small $\eta_{ij}$. Because of the presence of the strain gradient, mass density, i.e., specific volume, of the crystal is not uniform in space. Thus, a proper characteristic size or volume should be defined before discussing crystal instability. For an unstable crystal, instability is believed to begin from a lattice point whose fluctuation is the biggest. Then the instability develops rapidly and plastic process or phase transition begins. Thus, the characteristic size is at a scale of atom level. More detailed discussions could be found in *Appendix B*. With this physical picture in mind, we proceed by extending finite-strain continuum elasticity theory (Thurston and R., 1964; Wallace, 1970) and expanding free energy per characteristic volume ($\Omega$) as a function of $\eta_{ij}$ and $\kappa_{ijk}$ to the second order, that is

$$f_{\Omega_a}(\boldsymbol{\eta}(\boldsymbol{a}), \boldsymbol{\kappa}, T) = f_{\Omega_a}(0,0,T) + \sigma_{ij}\eta_{ij} + \tau_{lmn}\kappa_{lmn} + \frac{1}{2}C_{ijmn}\eta_{ij}\eta_{mn} + W_{ijlmn}\eta_{ij}\kappa_{lmn} +$$

$$\frac{1}{2}T_{ijklmn}\kappa_{ijk}\kappa_{lmn}, \tag{5}$$

where $f_{\Omega_a}(0,0,T)$ is the binding energy of the crystal at T, $\boldsymbol{\sigma}$ and $\mathbf{C}$ are stress tensor and elastic constants. Other quantities, conjugate to the strain gradient or its combinations, are defined by

$$\tau_{lmn} = \frac{1}{\Omega_a}\frac{\partial F}{\partial \kappa_{lmn}}\bigg|_{\boldsymbol{a}}, \tag{6}$$

$$W_{ijlmn} = \frac{1}{\Omega_a}\frac{\partial^2 F}{\partial \eta_{ij}\partial \kappa_{lmn}}\bigg|_{\boldsymbol{a}}, \tag{7}$$

and

$$T_{ijklmn} = \frac{1}{\Omega_a}\frac{\partial^2 F}{\partial \kappa_{ijk}\partial \kappa_{lmn}}\bigg|_{\boldsymbol{a}}. \tag{8}$$

Here we have use $F$ and $f$ to denote total free energy and average free energy density over volume $\Omega_{\mathbf{a}}$, respectively. To discuss stability of the strained crystal, we should consider variation of the free energy density in a small virtual strain gradient $\delta\widetilde{\boldsymbol{\kappa}}$ at configuration $\{X\}$. General derivations of the variation of the free energy density could be found in *Appendix A*. Under finite strain and strain gradient, the variation of the free energy density in $\delta\widetilde{\boldsymbol{\kappa}}$, as well as $\delta\widetilde{\boldsymbol{\eta}}$ caused by the strain gradient, is

$$\delta f_{V_X} = \left(\sigma_{IJ} + \tau_{KIJ,K}\right)\delta\widetilde{\eta}_{IJ} + \widetilde{B}_{IJMN}\widetilde{\eta}_{MN}\delta\widetilde{\eta}_{IJ} - \widetilde{T}_{LMNIJ}\widetilde{\kappa}_{LMN}\delta\widetilde{\eta}_{IJ}, \tag{9}$$

where $\delta\widetilde{\boldsymbol{\eta}}$ and $\widetilde{\boldsymbol{\eta}}$ are associated strain variation and strain measured with respective to $\{X\}$, and

$$\widetilde{B}_{IJKL} = C_{IJKL} + \frac{1}{2}\left(\sigma_{JL}\delta_{IK} + \sigma_{IL}\delta_{JK} + \sigma_{JK}\delta_{IL} + \sigma_{IK}\delta_{JL} - 2\sigma_{IJ}\delta_{KL}\right), \tag{10}$$

$$\tilde{T}_{LMNIJ} = T_{LMNKIJ,K}. \tag{11}$$

Higher than second-order terms of $\delta\tilde{\kappa}$ are neglected here. It should be noted that $\boldsymbol{\tau}$ and $\mathbf{W}$ are zero for centrosymmetric lattices under small strain gradient disturbance (See *Appendix B*), which could adequately reduce our calculations of $\tilde{\mathbf{T}}$. In present work, it is referred to as small strain gradient condition. According to (Fleck and Hutchinson, 1997; Mindlin and Eshel, 1968), work done by the high order stress $\boldsymbol{\tau}$ via a virtual strain gradient $\delta\tilde{\kappa}$ is $\tau_{ijk}\delta\tilde{\kappa}_{ijk}$ under linear strain gradient approximation. Then total work, done by the strain gradient and the associated strain, is

$$\delta W = (\sigma_{IJ} + \tau_{KIJ,K})\delta\tilde{\eta}_{IJ}. \tag{12}$$

The stability of a deform crystal at $\{X_i\}$ requires that the difference between the increase in free energy ($F$) and the work ($W$) done by the disturbance is positive, that is

$$V(\boldsymbol{X})(\delta F - \delta W) = \tilde{B}_{IJMN}\tilde{\eta}_{MN}\delta\tilde{\eta}_{IJ} - \tilde{T}_{LMNIJ}\tilde{\kappa}_{LMN}\delta\tilde{\eta}_{IJ} \geq 0. \tag{13}$$

For arbitrary $\tilde{\eta}_{MN}$, $\delta\tilde{\eta}_{IJ}$ and $\tilde{\kappa}_{LMN}$, this condition requires that $\tilde{\mathbf{B}}$ and $-\tilde{\mathbf{T}}^L$ (L = 1, 2, 3) are positive definite, where $\tilde{\mathbf{T}}^L$ is the L-th block matrix of $\tilde{\mathbf{T}}$. More details could be found in *Appendix A*. To further clarify this stability condition, we consider two special cases: 1) $\tilde{\boldsymbol{\kappa}} = \mathbf{0}$ and 2) $\tilde{\boldsymbol{\eta}} = \mathbf{0}$. For the first case, the stability condition only requires that $\tilde{\mathbf{B}}$ is positive definite, which is the modified B criteria derived from strain disturbance (Wang et al., 1995; Wang et al., 1993). For the second case, the stability condition is that $-\tilde{\mathbf{T}}^L$ (L = 1, 2, 3) are positive definite. Unlike the first case, the second case seems to be impossible in real word because any strain gradient would generate strains in crystals and leads to a violation of condition $\tilde{\boldsymbol{\eta}} = \mathbf{0}$. However, the stability condition (13) is developed in current deformation configuration, and for any strain gradient, there is at least a "point" where its strain is zero. Now, the second case is corresponding to the stability condition at the zero-strain "point" in the current configuration. Generally speaking, any strain and strain gradient states of a "point" in a crystal could be generated through applying a uniform strain followed by a strain gradient originating from this "point". Thereby, the stability condition of a deformed crystal with both strain and strain gradient is

$$\tilde{B}^e_{min} > 0 \;\; and \;\; \tilde{T}^e_{min}(L) > 0, \;\; (L = 1,2,3), \tag{14}$$

where $\tilde{B}^e_{min}$ and $\tilde{T}^e_{min}(L)$ are minimum eigenvalue of $\tilde{\mathbf{B}}$ and $-\tilde{\mathbf{T}}^L$. That is to say, the crystal would be instable when

$$\tilde{B}^e_{min} < 0 \;\; or \;\; \tilde{T}^e_{min}(L) < 0, \;\; (L = 1,2,3). \tag{15}$$

It is convenient to contract the four indexes of $\tilde{\mathbf{T}}^L$ ($\tilde{\mathbf{B}}$) into two by Voigt notation. Obviously, $\tilde{\mathbf{T}}^L$ ($\tilde{\mathbf{B}}$) is a symmetric matrix which has six real eigenvalues. Thus, the condition (15) could be generally applied for judging the instability of any deformed crystals. Interestingly, the condition (15) consists of two subconditions which are corresponding to the two special cases mentioned above. In remaining parts of present work, we refer to the first subcondition, corresponding to the first case, as strain instability criteria and the second as SGI criteria. The instabilities of a crystal are a competing result between strain instabilities and gradient instabilities.

Next, we will investigate the role of strain gradients played on the instabilities based on the gradient stability criteria. For the ramp compressions along +Z direction of single crystalline iron, only uniaxial compressions with uniaxial gradients along Z direction are considered here. Assuming that uniaxial compression ratio of current configuration is λ, the only non-zero component of linear strains and finite Lagrangian strains are $\varepsilon_{33}$ = λ-1 and $\eta_{33}$ = (λ$^2$-1)/2, respectively. Without ambiguity, we omit the subscript of the uniaxial gradient $\kappa_{333}$ (and $\eta_{33}$) in the remaining text. Then we could calculate $\tilde{\mathbf{T}}^L$ matrix at different strains, by using expression (B.37) derived in *Appendix B*, which has made use of the small strain gradient condition. Because $\tilde{\mathbf{T}}^1 = \tilde{\mathbf{T}}^2$, we only need to check the minimum eigenvalue of $-\tilde{\mathbf{T}}^1$ and $-\tilde{\mathbf{T}}^3$. The minimum eigenvalue of $-\tilde{\mathbf{T}}^L$ calculated at different strain and strain gradient are shown Fig. 5c and 5d. It is found that the minimum eigenvalue of $-\tilde{\mathbf{T}}^1$ first become negative under uniaxial compressions with small strain gradient disturbance. This indicates that the instabilities begins in transverse (X or Y) directions rather than the compression direction, which is well consistent with our observations (See S4). Besides, critical strain, at which $T_{\min}^1$ begins to be negative, is 0.086 which is indeed the critical strain observed in the ramp compressions. Thus, we could conclude that the instability under ramp compressions takes place due to the disturbance of uniaxial strain gradient rather than uniaxial strain, and it is triggered by transverse strain gradient instability. As shown in Fig. 5c, the predicted critical strain by the SGI criteria gradually deviates from the ones observed in the ramp compressions when the strain gradient is larger than 1.43e-4 Å$^{-1}$, corresponding to the strain rates larger than 0.053 Å/ps$^2$ in our simulations. This is because expression (B.37) is derived for small strain gradient. To obtain a more precise prediction value at large strain gradient, the instability condition should be evaluated under finite strain gradient where both $\boldsymbol{\tau}$ and $\mathbf{W}$ would contribute to the instability. This is a routine task by methods proposed in *Appendix B*, but further investigations of the instability are beyond scopes of present studies. As a conclusion of the above discussions, we have proved that dynamic critical strain (or yield limit) is different from the static one due to the presence of strain gradient disturbances.

The onsets of the phase transition at different $\dot{\varepsilon}$ are plotted in Fig. 8, where strain rate dependence of the TP obeys a power law when $\dot{\varepsilon}C > 0.053$ Å/ps$^2$ (about $10^{10}$ s$^{-1}$). Similar power law is only observed in experiments (Smith et al., 2013), which is attributed to thermal activation mechanism similar to that of plastic deformations of metals (Smith et al., 2011). However, the mechanism does not work here because temperature keeps at its initial value (0 K) until the phase transition takes place (See Fig. 1). In present work, the strain rate dependence of the TP is due to the time needed to nucleate from bcc phase to an hcp nucleus. Under static compressions at 0K, single crystalline iron will begin to nucleate after the bcc phase is compressed to a critical strain, above which bcc iron is instable. While under high strain rates, the bcc phase does not have enough nucleation time ($t_0$) to finish the transition from a bcc phase to an hcp nucleus so that it is carried into a higher stress state than the onset of instabilities. This means that the excess of strain ($\Delta\varepsilon$) over the critical strain ($\varepsilon_c$) of the instability increases with growth of the applied strain rates by $t_0\dot{\varepsilon}$, which lead to a strain-rate-dependence TP. Considering relaxations of local stresses (and thus strain) due to the instability, the strain-rate-dependence TP may be expressed by $\boldsymbol{\sigma}_{TP}(\varepsilon_c' + t_0\dot{\varepsilon})$, where $\varepsilon_c'$ is the relaxed critical strain. The nucleation time is approximately proportional to $\upsilon^{-1}$, where $\upsilon$ is optical phonon frequency of compressed bcc cell at the onset of the instability. According to the stability analyses, the critical strain (and thus $\upsilon$) is nearly a constant when the applied strain rate is smaller than 0.053 Å/ps$^2$. This is because higher order stress, associated with the strain gradient, is zero within the strain rate range. However, the higher order stress ($\boldsymbol{\tau}$) could not be neglected when the applied strain rate is larger than 0.053 Å/ps$^2$. In this case, the TP should be evaluated by $\boldsymbol{\sigma}_{TP}(\varepsilon_c' + t_0\dot{\varepsilon}) + \nabla \cdot \boldsymbol{\tau}$ which will leads to a different dependence relationship between TP and strain rate. As shown in Fig. 7, systematical deviations from the linearly fitting are emerged when $\dot{\varepsilon}C > 0.053$ Å/ps$^2$ which is just corresponding to the condition when the higher order stress cannot be neglected. From the discussion in this part, we find that applied strain rate with a value smaller than 0.053 Å/ps$^2$ would only provide a small strain gradient disturbance in the compressed sample, and thus affect the lattice instabilities. The strain rate effects on the TP are reflected via the nucleation time of the phase transition. However, a higher applied strain rate would generate a finite strain gradient effect (a non-zero higher order stress and work conjugates) which could apparently affect onsets of the instabilities and the TP.

*3.3 Interactions between Compression Wave and Nucleation of the Phase Transition*

According to our simulation results shown in Fig. 1-2, phase transition begins to nucleate at the instability region only when disturbance fields are present. The disturbance may be temperature fluctuations or additional strain filed generated by newly forming lattice defects. Because the instability does not cause an apparent rising in temperature, the temperature of the instability region keeps the same as that of the initial sample, i. e., 0K in present work. Thus, temperature fluctuations are not the source of the disturbance in this case. We believe that the disturbances of a nucleation site are caused by strain fields generated by newly forming nucleuses nearby. And the generated strain fields will trigger a next nucleation nearby. These processes are repeated as more and more nucleuses nucleate from the initial nucleation site to its surrounding in the instability region. Then the nucleation quickly fills the transverse dimension of the instability region and creates a macroscopic moving boundary of nucleation along Z direction. According to this physical picture, the speed of the moving boundary of nucleation could be evaluated as follows. A schematic drawing of the nucleation of hcp phase has been shown in Fig. 8, where the gray and black bars denote adjacent shuffle planes, i.e., (110) or (1-10), in bcc iron. Nucleation could begin by consecutively shuffling layer by layer or simultaneously shuffling several layers at one time. Bertrand et. al. (Dupé et al., 2013) compared the two transition path by *ab initio* calculations and found that the consecutive shuffle manner is more energetically favorable. However, more complex consecutive shuffle manners are not explored, for example, nucleation proceeds by shuffling two or more layers each time. Here, we assume the number of active layers is *m* during the nucleation. Every shuffle will move the phase interface forwards a distance of 2*md* along [110] (or [1-10]) direction, where *d* is the distance between two adjacent $\{110\}$ planes, that is $\sqrt{2}a_0/2$, $a_0$ is lattice constant of bcc iron. Since the nucleation proceed in the instability region where nearly no energy barrier is needed to overcome during every try of the shuffle by atom layers, the try frequency of the shuffle is equal to optical phonon frequency (*v*) along [110] or [110] direction in reciprocal space of compressed bcc lattice. Thus the moving speed of the phase interface along [110] (or [1-10]) is 2*mdv*. Because phase transition domain is always close to be ellipsoidal under homogeneous nucleation mechanism, the moving speed of the phase interface along the wave propagation direction (or [001] direction) is proportional to the speed along [110] (or [1-10]). Let χ$_0$ to be the ratio between the principal-axis lengths of the ellipsoidal along [110] (or [1-10]) and [001] at unstrained state (See Fig. 9), the ratio at a compressed state, whose uniaxial compression ratio is λ, is λχ$_0$ for first layer. Considering existence of a small uniaxial strain gradient ($\kappa$), the uniaxial compression ratio at layer *i* relates to the one at layer *0* by $\lambda_i = \lambda_0 + id\kappa$. Thus the moving speed of the phase interface along [001] is

$$U_{||} = 2\chi_0 dv \sum_{i=0}^{m-1}(\lambda_0 + id\kappa) + v_p, \tag{16}$$

where $v_p$, the mass center velocity of the nucleus, could be approximately evaluated by the average particle velocity at the nucleation position. Because movements of the boundary of the nucleation region are caused by the inter-triggering between nucleation and strain field disturbance, the resulting boundary speed is sum of the moving speed of the phase interface of new nucleuses and the propagation speed of strain field generated by the new nucleuses, that is

$$U_B = U_\parallel + c = 2\chi_0 dv \sum_{i=0}^{m-1}(\lambda_0 + id\kappa) + v_p + c, \tag{17}$$

where *c* is the propagation speed of strain field.

### *3.4 Nucleation of the Phase Transition after Formation of the Initial Shock*

The time evolutions of strain ahead of the IB are shown in Fig. 10, where the strain gradually decreases during the wave propagations until the initial shock forms. And absolute value of finial strain at front of the instability region grows with the increasing of the strain rate. As discussed in former sections, the initial instability is induced by mechanical instability of bcc iron, which is much stricter than equilibrium thermodynamic conditions of transforming from bcc to hcp phase. In order to form a nucleus of the hcp phase, the Gibbs free energy of compressed bcc phase should be larger than that of hcp phase due to nucleation barrier and additional strain energies introduced by the new nucleus. Considering the time required by the strain field of the new nucleus to propagate, the strain energies are not a key factor in determining driven force of the phase transition under ultrafast compressions. If applied compression continuingly acts on a nucleus at a rate faster than growth speed of strain energy induced by the new nucleus, the nucleus would be able to grow if the driven force is large enough to overcome the nucleation barrier. Otherwise, it may disappear at the wave front of the instability region. Nevertheless, nucleation could take place under the continuing compressions for enough time. The nucleation barrier could be evaluated by transition states in the transformation path from bcc to hcp phase. To achieve a transformation from bcc to hcp phase, the energy of the compressed bcc phase should be large enough to enable a subsequence shuffle processes among {110} planes. Obviously, the energy of compressed bcc phase at the nucleation site is larger than its product phase whose initial strain is dependent on the compression degree of the bcc phase. In other words, the bcc phases of different strains at the front of the instability region will transform into hcp phases with different initial strains. For example, the ideal transition path requires a compression ratio of 86.6% along [001] direction of initio bcc phase before the shuffle processes among (110) (or (1-10)) planes, which generates a strain-free hcp phase when neglecting the slight expansion along [110] (or [1-10] direction). However, the shuffle processes could probably happens in a compressed bcc phase with a compression ratio less than 86.6%, and thus generates a strained hcp phase whose energy is still lower than the compressed bcc phase. Similarly, the shuffle processes may also do not proceed ideally, which will left a shear along the shuffle planes. Though the strained hcp phase is not as stable as an ideal hcp phase in formation energy, the misfit energy between the strained hcp nucleus and the surrounding compressed bcc phase is smaller than the one between the ideal hcp nucleus and the bcc phase. If the sum of the formation energy of the product phase and the misfit energy between the product phase and the surrounding parent phase reach to a minimum, the transition path could take place and result in a residual strain in the phase products. The detailed transition path relies on the states (mainly including strain and strain gradient here) of compressed parent phases. *Ab*-initio calculations on the transition path of the phase transition of iron also uncovered the possibility of different transition paths under different pressures albeit with the ideal processes for both the compression and shuffling (Lu et al., 2014). In contrast to ideal transition path, it may be called "strained transition path". The strain rate dependence of the strain ahead of the IB (See Fig. 10) could be explained by the strained transition path. According to the strained transition path, we could infer that c/a ratio, closely related to the strain of the transition product after the transition, also depends on the strain rate, which is consistent with a recent experimental analyses (Konôpková et al., 2015).

From the physical picture of the wave propagations during ramp compressions, we know that a steady shock wave could form when the NB catches up to the IB. Let $c_0$ to be the speed of the IB, then the steady wave condition is $U_B = c_0$. According to equation (17), we have

$$v_p = -2\chi_0 d\nu\lambda_0 + c_0 - c. \tag{18}$$

In the above equation, we have taken $m$ to be 1 because nucleation region, at the steady shock wave front, is too narrow to allow more number of shuffle processes to take place simultaneously. Equation (18) indicates that the compression ratio is linearly related to the particle velocity just ahead of the IB after a steady shock wave forms. As shown in Fig. 11, the relationship between the compression ratio (or linear strain ε) and the particle velocity of bcc phase does indeed satisfy equation (18). Assuming that Lagrangian wave speed of the IB is $C_L$, the relation is equivalent to the simple wave relation, i.e., $dv_p = -C_L d\varepsilon_0$, if $C_L = 2\chi_0 d\nu$. This is valid if effects of strain rates and dissipative processes, for example, plasticity and phase transition, need not to be considered in the equation of states. That is to say, before the instability takes place, an inviscid scalar equation of states is a good approximation for describing single crystalline iron under ramp compressions at low temperature. This is nontrivial because the strain rate effects cannot be ignored at the front of the initial shock. Namely, compared with elastic response of crystals, elastic instability is more sensitive to the strain gradient. This result is reasonable since the elastic instability depends on the disturbance (very small strain gradients would trigger the instability as long as the instability criteria are satisfied), while the elastic response relies on the magnitude of strain gradients which are usually much smaller than that of strain. The linear relationship (18) indicates that $\nu\chi_0$ is nearly a constant under ramp compressions. Through fitting to equation (16), we find that $c \sim c_0 + C_L$, and $\nu\chi_0 \approx 12$ THz. The vibrational frequency of (110) planes, corresponding to the zone-boundary transverse optical phonon mode, could be estimated through compressing bcc iron along [001] direction to different compression ratios and performing lattice dynamic calculations with the potential employ in this work. As shown in Fig. 12, $\nu$ lies within ranges of [9.2, 10.8] THz for λ∈[-0.91, 1.0]. And the corresponding variations of $\chi_0$ is within ranges of [1.1, 1.3].

As shown in Fig. 10, the formation time ($t_f$) of the initial shock could be approximately evaluated by the scaling method proposed by Lane and et. al. (Lane et al., 2016), that is

$$t_f = t' \frac{v'_{max} t_{rising}}{v_{max} t'_{rising}} = t' \dot{\varepsilon}'/\dot{\varepsilon}, \tag{19}$$

where quantities with a upper prime is corresponding to the ones measured in a reference system. Here, we take the reference system to be the simulated iron sample under a ramp compression with $v'_{max}$ = 0.8 km/s and $t'_{rising}$ = 5 ps, and thus $t'$, the formation time of the initial shock, is about 16ps. Via using equation (19), we could infer that the formation time for the ramp compression with $v_{max}$ = 1.5 km/s and $t_{rising}$ = 50 ps is 85 ps which is larger than the max allowed wave propagation time in present work (See Fig. 10). However, we find that the scaling law (19) is also violated when the strain rate is larger than $10^{10}$ s$^{-1}$ (See Fig. 10 and S5). This is also due to the finite strain gradient effect.

## 4. Summary and Conclusions

In summary, ramp compressions upon single crystalline iron have been conducted to study the phase transition and the formation process of a steady shock wave through atomic simulations, where applied strain rate ranges from $10^9$ to $10^{11}$ s$^{-1}$. Strain rate dependence of the TP obeys a power law when the strain rate is smaller than $10^{10}$ s$^{-1}$, which is consistent with experiments conducted by Smith ant et. al (Smith et al., 2013). Nucleation of transition products begins after bcc iron is compressed to an unstable state, characterized by a critical strain. The critical strain is observed to be about 0.086 and would gradually decrease with the increment of the applied strain rates. A thorough theoretical analysis on the critical strain is made through comparing observed value from the NEMD simulations with the one predicted by lattice dynamic methods. We find that the widely used modified B criteria could neither correctly predict the observed critical strain nor explain the rate dependence of the critical strain. Since the modified B criteria have successfully predicted critical strains of many crystals under quasi-/static compressions, its failure on prediction of the dynamic critical strain result in an essential theoretical difficulty in understanding lattice instabilities under dynamic loadings. By comparing these results with the ones obtained under homogeneous uniaxial compressions, strain gradient is found to be a key factor to discriminate dynamic compressions from the static ones. Through considering contributions of the strain gradient, as well as strain, on the instabilities, new instability criteria are developed in a general form. With the new instability criteria, both static and dynamic instabilities could be consistently explained. Especially, the critical strain predicted by the new instability criteria is demonstrated to be the same with the one observed from NEMD simulations under small strain gradient disturbance. However, the observed critical strains deviate from the predicted value when applied strain rate is larger than $10^{10}$ s$^{-1}$. The reason is that contributions of higher order stresses and work conjugates of strain gradient on free energies, referred to as finite strain gradient effect, cannot be neglected under extreme strain rates. The finite strain gradient effect is also found to affect the TP and the scaling law obeyed by formation time of the initial shock when the applied strain rate is larger than $10^{10}$ s$^{-1}$.

Besides, formation process of a steady shock wave under ramp compressions is found to be made up of three events sequentially: lattice instability begins and propagates; nucleation takes place in the instability region and proceeds towards to the IB ahead of it; phase interface between the completely transitioned region and the nucleation region moves towards the NB. The moving speed of the three boundaries increases sequentially in order of the IB, the NB and the phase interface, which will eventually lead to the formation of a single steady shock wave. Before the formation of the single steady shock wave, an initial shock may be formed in advance when the IB is caught up by the NB. And the moving speed of the NB after the formation of the initial shock could be well explained by an inter-triggering mechanism between the nucleation and the strain field of new nucleuses. The big rinsing in temperature is mainly caused by the moving phase interface which contributes to the formation of the finial single shock wave. Based on the reason of the movement of the NB and the formation condition of the initial shock, a linear relation between particle velocity and strain at the position ahead of the shock front is established. We find that the linear relation is generally applied for the whole ranges of strain rates involved in this study. Through comparing the linear relation with the simple wave relations, we find that the simple wave assumption is valid until bcc iron is compressed to an instable state. This result indicates that an inviscid scalar equation of states is a good approximation for describing single crystalline iron under ramp compressions before instabilities takes place.

# Acknowledgements

This work is supported by the National Natural Science Foundation of China (NSFC-NSAF 11076012 and NSFC 11102194, 11402243), National Key Laboratory Project of Shock Wave and Detonation Physics (No. 077120), the Science and Technology Foundation of National Key Laboratory of Shock Wave and Detonation Physics (Nos. 9140C670201110C6704 and 9140C6702011103) and Chinese National Fusion Project for ITER with Grant No. 2013GB114001.

Fig. 1. Profiles represented by (a) temperature, (b) particle velocity, (c) nearest neighbor separation and (d) coordination number of iron samples under ramp compression along [001] direction with a max particle velocity of 0.8 km/s and a ramp rising time of 15ps. The first kink in each profile (marked by a downward red triangle above each profiles) is caused by the transition from bcc phase to the TCN structure, which indicates the onset of lattice instability. Phase transition proceeds in the unstable region. Positions of the NB at each moment are marked by an upward blue triangle below the corresponding profiles.

Fig. 2. Phase evolutions of the simulated iron sample during ramp compression with an applied strain rate of 8 /15 Å/ps$^2$. All snapshots of the simulated sample at each moment are colored by coordination number of atoms: dark blue (12), light blue (11), green (10), yellow (9), and red (8), where compression waves propagate from left to right. The instability interface (or phase interface) is marked by a gray (black) bar in each snapshot. The unstable region disappears from the wave profiles when the phase interface catch up to the instability interface. A mixed-phase region emerges before a complete transition from the bcc phase to the hcp phase.

Fig. 3. Time evolutions of wave profile, represented by potential energy per atom, under ramp compressions with a strain rate of 8/5 Å/ps$^2$, where a steady single shock wave could be observed after 48ps.

Fig. 4. Spatial distributions of (a) strain and (b) strain gradient for the strain rates listed in the figure at 4, 8, 16and 50 ps, respectively. These times correspond to the moments when the compressed bcc iron begins to become instable. For each applied strain rate, the distribution curve begins from the position (marked by red balls in (a)) where the instability takes place. In response to the applied strain rates, the strain of the bcc iron increases linearly to the "A" point and then nonlinearly to the critical strain of the instability. The inset in the figure (b) is corresponding to the strain gradient distribution of the compressed bcc iron under ramp compressions with a strain rate of 8/10 Å/ps$^2$, where the red line is used as a guiding for the eye.

Fig. 5. a) Strain, as well as strain gradient, at the onset of instabilities during ramp compressions with different applied strain rate. The black line is a guiding for eyes, which forms a boundary departing stable region from instable region. The horizontal dashed line is corresponding to a critical strain of -0.086 at small strain gradient, while the vertical dot-dashed line marks a critical strain gradient, beyond which finite strain gradient effect cannot be neglected. b) Minimum eigenvalue of elastic constants of bcc iron as a function of uniaxial strain. The critical strain (marked by "B") is predicted to be -0.128 under uniaxial compressions. c) $\tilde{T}^1_{min}$ and d) $\tilde{T}^3_{min}$ as a plot of uniaxial strain and uniaxial strain gradient for single crystalline iron, where the unit is GPa·Å$^2$. Dynamic instability point, observed in ramp compressions, has been marked by "A" in all of the figures.

Fig. 6. Pressure versus strain under homogeneous uniaxial compressions along [001] direction of perfect iron single crystal with different strain rates, where the black arrows have marked the onsets of instabilities. When strain rates are smaller than $10^{10}$ s$^{-1}$, critical strain of the instabilities is about 0.11 which is close to the value predicted by the modified B criteria. However, the critical strain grows with the increasing of strain rate when the strain rate is larger than $10^{10}$ s$^{-1}$.

Fig. 7. The transition pressure of iron as a function of uniaxial strain rate, represented by log $\dot{\varepsilon}$. The gray dashed line is a fitting of a power law of $\sigma_{TP} = A_0 \dot{\varepsilon}^n$, where $A_0 = 0.326$, $n = 0.196$. Obviously, the fitting over whole ranges of strain rates is not good, but if we omit the points larger than $10^{10}$ s$^{-1}$, a good fitting to the power law would be achieved (as marked by red solid line). Deviation from the power law, at a strain rate larger than $10^{10}$ s$^{-1}$, results from the finite strain gradient effect where higher order stresses become increasing important. The deviation could also be observed from experiments performed by (Smith et al., 2013) and (Amadou et al., 2016) albeit with their different turning points due to different iron samples.

Fig. 8. Schematic drawing of propagations of phase interface and instability interface during ramp compressions, whose reference frame moves at a speed equal to that of the instability interface. In considering the shuffle planes of the phase transition for the shock along [001] direction, the long bars in the sample denote (110) (or (1$\bar{1}$0)) planes of bcc iron. The transition wave is assumed to propagate along [110] (or [1$\bar{1}$0]) direction which is normal to the [001] direction. The region nearby the phase interface in (b), circled by the dashed square, represents a spatial range affected by the phase interface. If the spatial range is doubled, the evolution sequence may be (a)→(c) rather than (a)→(b) →(c). Different spatial range will create a different sequence. However, whatever spatial range it is, the active atom layer is one (as circled by the dash square) when the phase interface catch up to the instability interface. This is because shuffle processes can proceed only in the unstable region.

Fig. 9. (a) Nucleation of hcp phase at 16ps under ramp compressions with $v_{max}$ = 0.8 km/s and t$_{rising}$ = 15ps and (b) a schematic drawing of an hcp nucleus in (a) as marked by the blue arrow.

Fig. 10. Time evolution of the strain ahead of the IB. The blue arrows mark the shock-formation time predicted by Eq. (17).

Fig. 11. Particle velocity as a function of strain ahead of the front of instability region for different ramp compressions performed in this work, where the strains, at the moment when the first shock forms, are marked by the red crosses. The red line is a linear fitting to the marked strains, whose fitting expression has been shown in the figure.

Fig. 12. Phonon dispersion curves of bcc iron under uniaxial compressions with different compression ratio, where the transverse optical phonons at the zone boundary are marked by red balls.

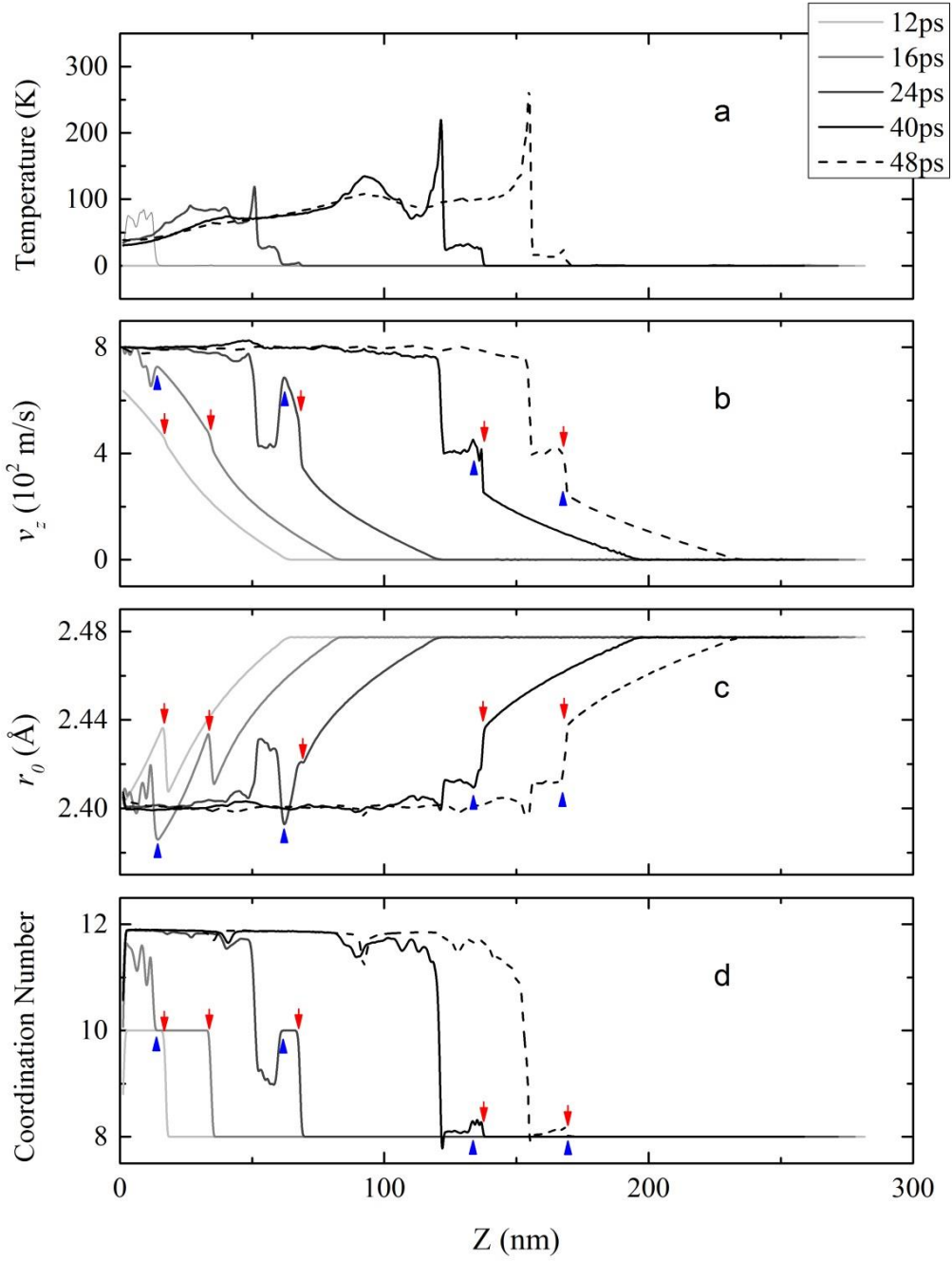

Fig. 1. Profiles represented by (a) temperature, (b) particle velocity, (c) nearest neighbor separation and (d) coordination number of iron samples under ramp compression along [001] direction with a max particle velocity of 0.8 km/s and a ramp rising time of 15ps. The first kink in each profile (marked by a downward red triangle above each profiles) is caused by the transition from bcc phase to the TCN structure, which indicates the onset of lattice instability. Phase transition proceeds in the unstable region. Positions of the NB at each moment are marked by an upward blue triangle below the corresponding profiles.

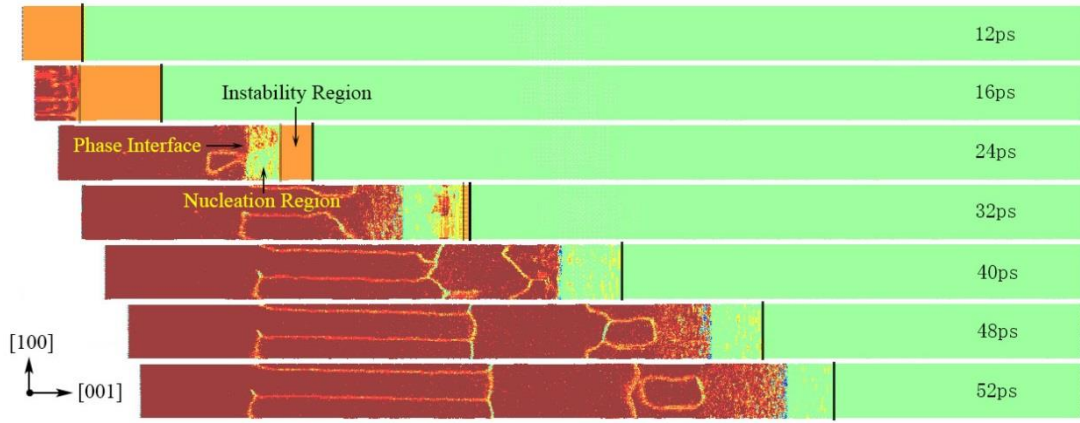

Fig. 2. Phase evolutions of the simulated iron sample during ramp compression with an applied strain rate of 8 /15 Å/ps$^2$. All snapshots of the simulated sample at each moment are colored by coordination number of atoms: dark blue (12), light blue (11), green (10), yellow (9), and red (8), where compression waves propagate from left to right. The instability interface (or phase interface) is marked by a gray (black) bar in each snapshot. The unstable region disappears from the wave profiles when the phase interface catch up to the instability interface. A mixed-phase region emerges before a complete transition from the bcc phase to the hcp phase.

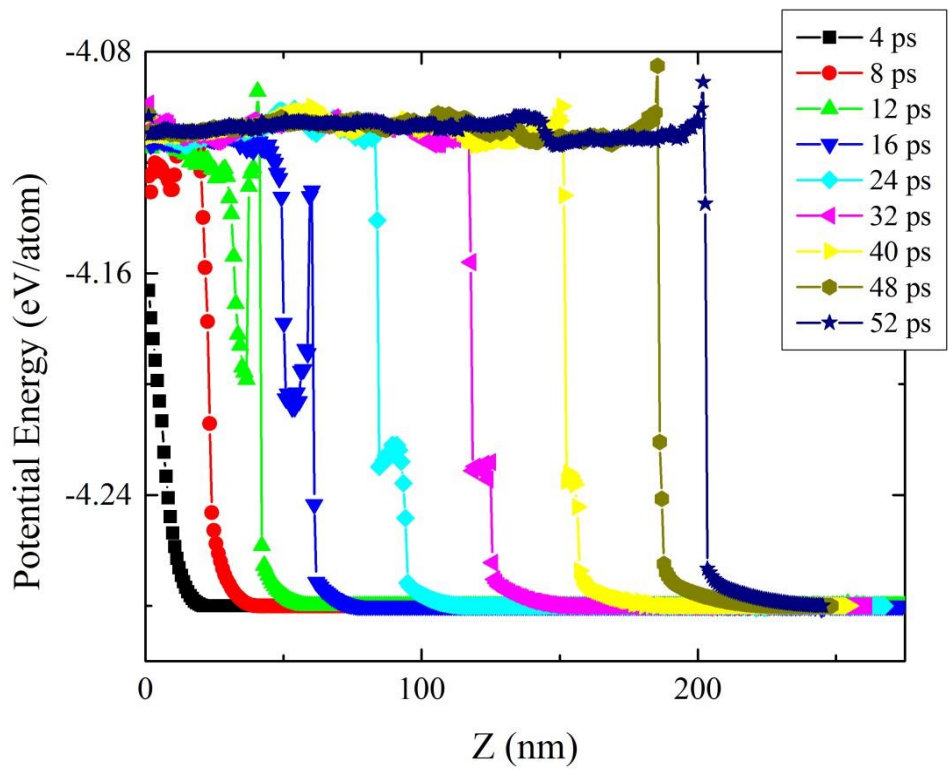

Fig. 3. Time evolutions of wave profile, represented by potential energy per atom, under ramp compressions with a strain rate of 8/5 Å/ps$^2$, where a steady single shock wave could be observed after 48ps.

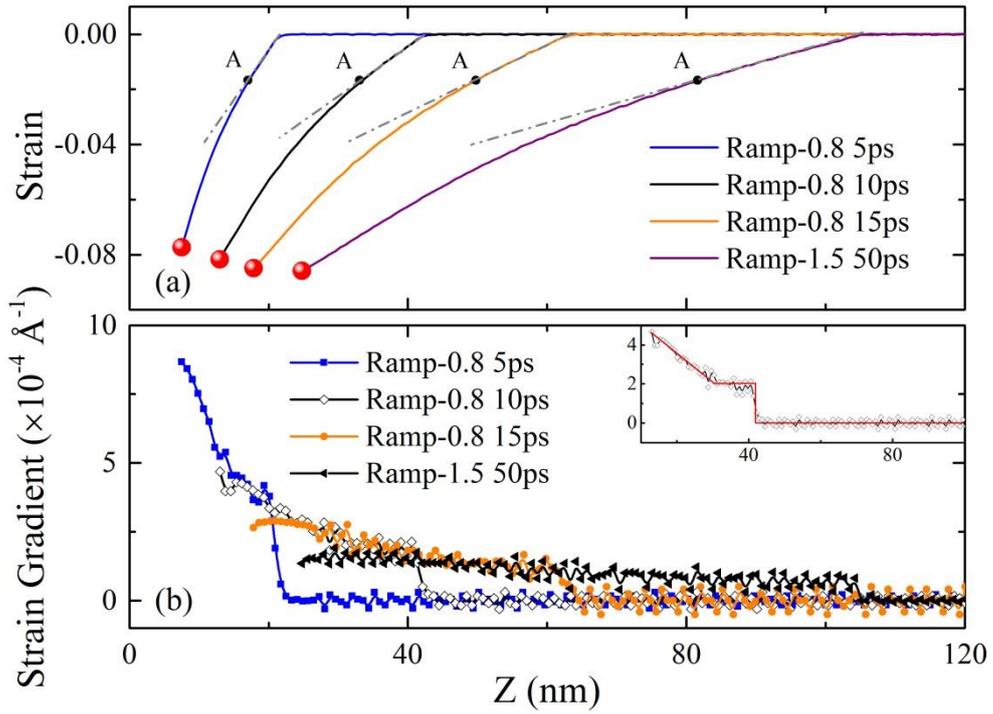

Fig. 4. Spatial distributions of (a) strain and (b) strain gradient for the strain rates listed in the figure at 4, 8, 16 and 50 ps, respectively. These times correspond to the moments when the compressed bcc iron begins to become instable. For each applied strain rate, the distribution curve begins from the position (marked by red balls in (a)) where the instability takes place. In response to the applied strain rates, the strain of the bcc iron increases linearly to the "A" point and then nonlinearly to the critical strain of the instability. The inset in the figure (b) is corresponding to the strain gradient distribution of the compressed bcc iron under ramp compressions with a strain rate of 8/10 Å/ps$^2$, where the red line is used as a guiding for the eye.

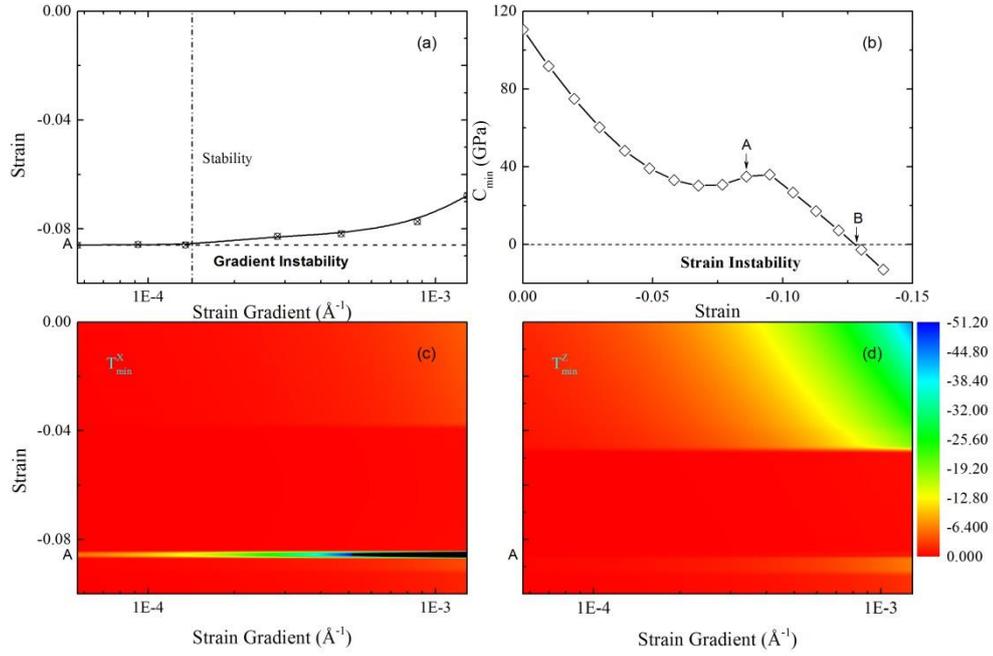

Fig. 5. a) Strain, as well as strain gradient, at the onset of instabilities during ramp compressions with different applied strain rate. The black line is a guiding for eyes, which forms a boundary departing stable region from instable region. The horizontal dashed line is corresponding to a critical strain of -0.086 at small strain gradient, while the vertical dot-dashed line marks a critical strain gradient, beyond which finite strain gradient effect cannot be neglected. b) Minimum eigenvalue of elastic constants of bcc iron as a function of uniaxial strain. The critical strain (marked by "B") is predicted to be -0.128 under uniaxial compressions. c) $\widetilde{T}^1_{min}$ and d) $\widetilde{T}^3_{min}$ as a plot of uniaxial strain and uniaxial strain gradient for single crystalline iron, where the unit is GPa·Å$^2$. Dynamic instability point, observed in ramp compressions, has been marked by "A" in all of the figures.

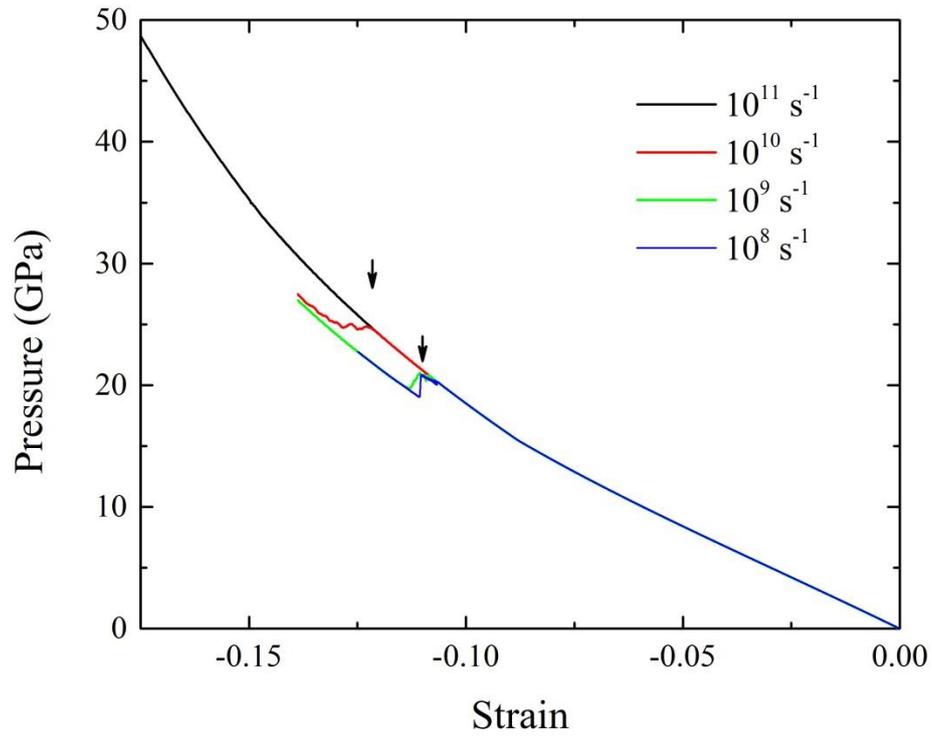

Fig. 6. Pressure versus strain under homogeneous uniaxial compressions along [001] direction of perfect iron single crystal with different strain rates, where the black arrows have marked the onsets of instabilities. When strain rates are smaller than $10^{10}$ s$^{-1}$, critical strain of the instabilities is about 0.11 which is close to the value predicted by the modified B criteria. However, the critical strain grows with the increasing of strain rate when the strain rate is larger than $10^{10}$ s$^{-1}$.

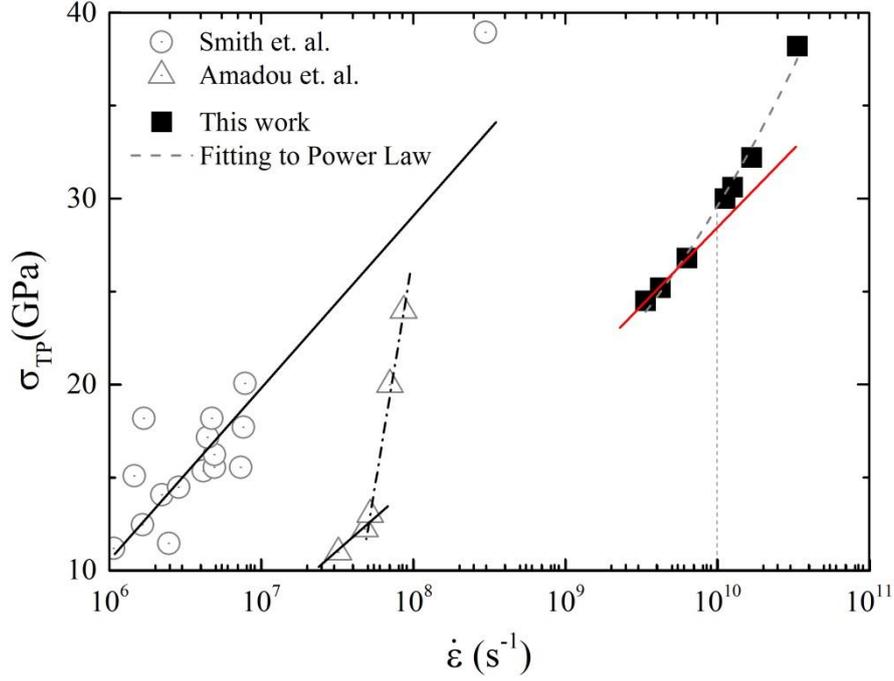

Fig. 7. The transition pressure of iron as a function of uniaxial strain rate, represented by log $\dot{\varepsilon}$. The gray dashed line is a fitting of a power law of $\sigma_{TP} = A_0\dot{\varepsilon}^n$, where $A_0 = 0.326$, $n = 0.196$. Obviously, the fitting over whole ranges of strain rates is not good, but if we omit the points larger than $10^{10}$ s$^{-1}$, a good fitting to the power law would be achieved (as marked by red solid line). Deviation from the power law, at a strain rate larger than $10^{10}$ s$^{-1}$, results from the finite strain gradient effect where higher order stresses become increasing important. The deviation could also be observed from experiments performed by (Smith et al., 2013) and (Amadou et al., 2016) albeit with their different turning points due to different iron samples.

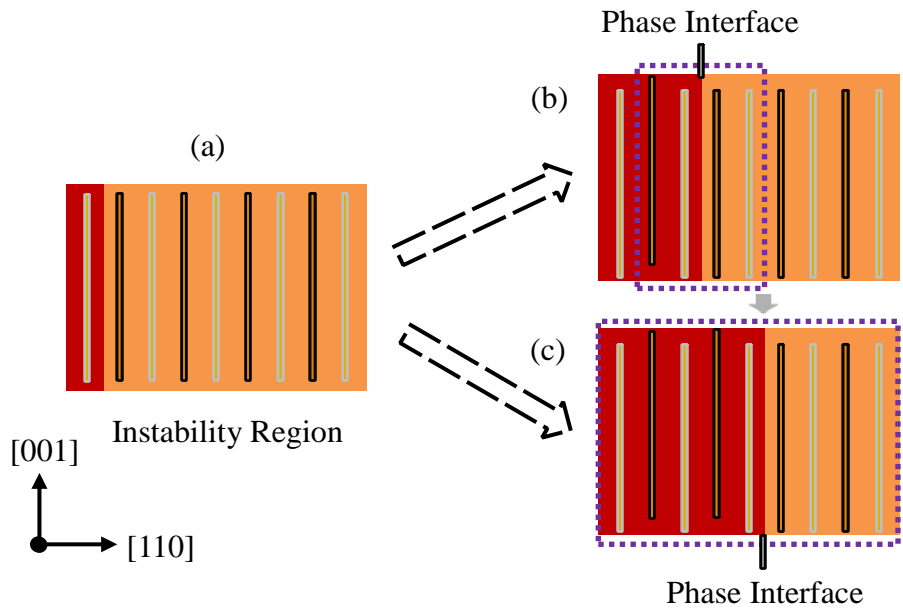

Fig. 8. Schematic drawing of propagations of phase interface and instability interface during ramp compressions, whose reference frame moves at a speed equal to that of the instability interface. In considering the shuffle planes of the phase transition for the shock along [001] direction, the long bars in the sample denote (110) (or ($1\bar{1}0$)) planes of bcc iron. The transition wave is assumed to propagate along [110] (or [$1\bar{1}0$]) direction which is normal to the [001] direction. The region nearby the phase interface in (b), circled by the dashed square, represents a spatial range affected by the phase interface. If the spatial range is doubled, the evolution sequence may be (a)→(c) rather than (a)→(b) →(c). Different spatial range will create a different sequence. However, whatever spatial range it is, the active atom layer is one (as circled by the dash square) when the phase interface catch up to the instability interface. This is because shuffle processes can proceed only in the unstable region.

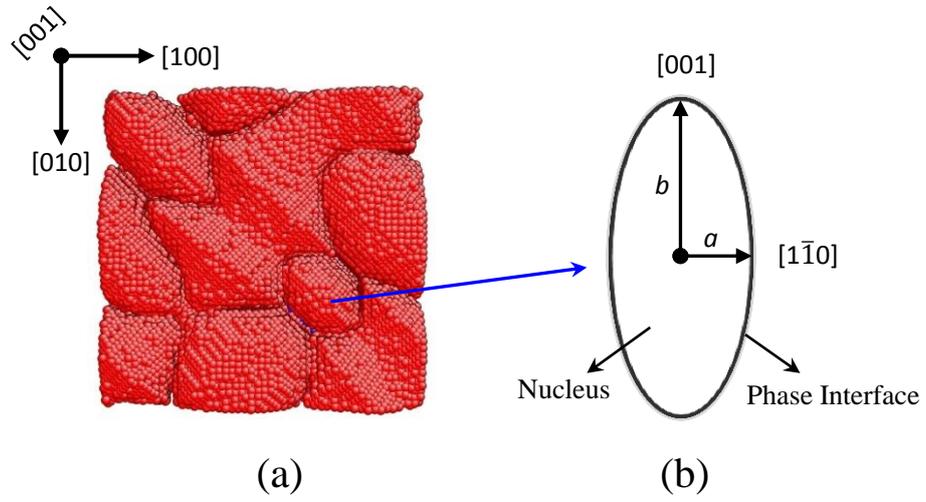

Fig. 9. (a) Nucleation of hcp phase at 16ps under ramp compressions with $v_{max}$ = 0.8 km/s and $t_{rising}$ = 15ps and (b) a schematic drawing of an hcp nucleus in (a) as marked by the blue arrow.

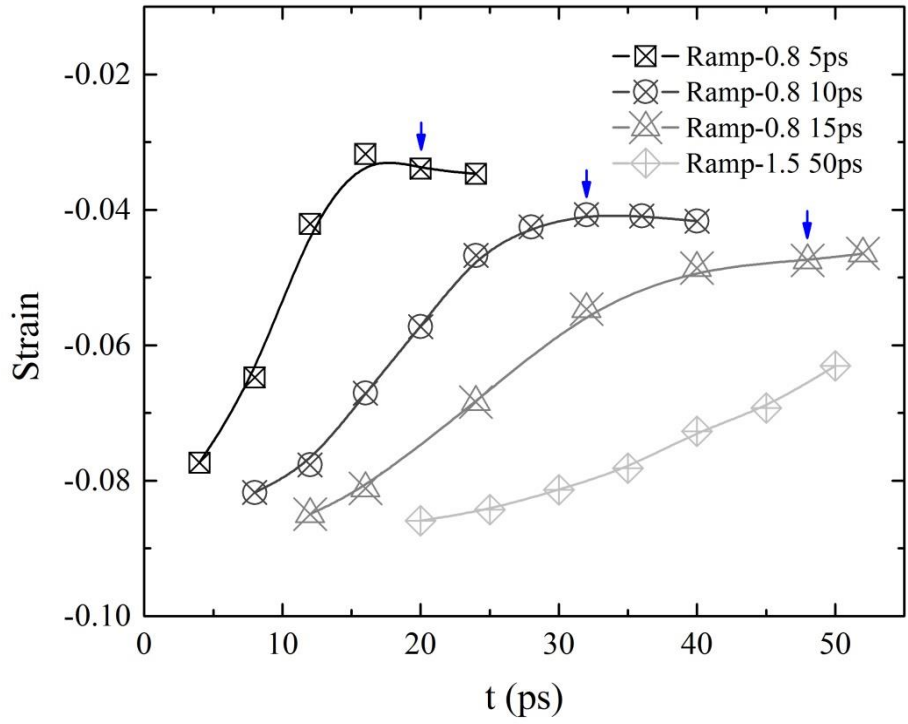

Fig. 10. Time evolution of the strain ahead of the IB. The blue arrows mark the shock-formation time predicted by Eq. (17).

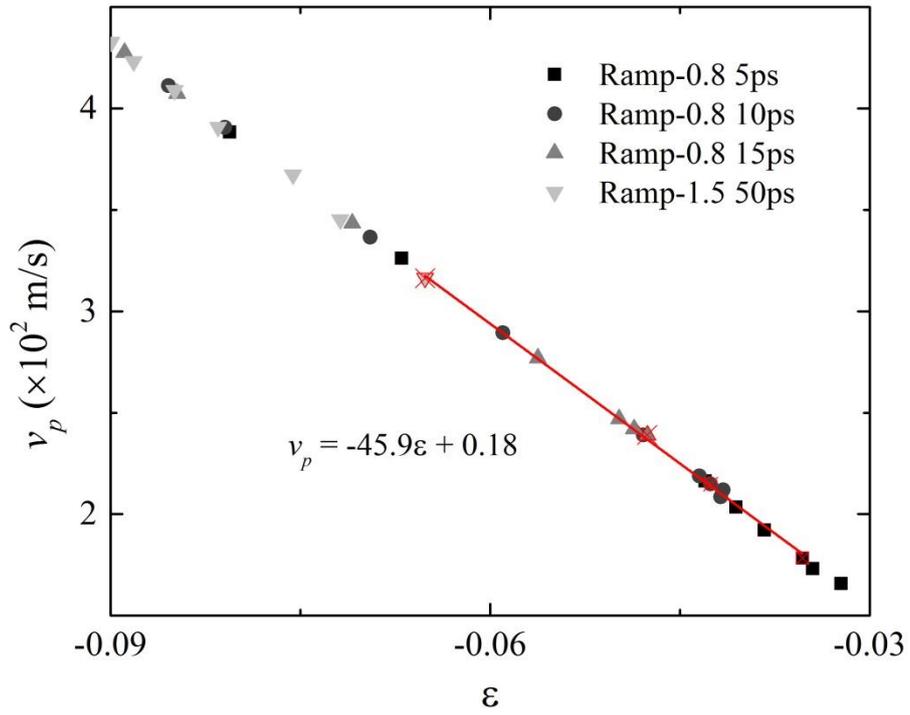

Fig. 11. Particle velocity as a function of strain ahead of the front of instability region for different ramp compressions performed in this work, where the strains, at the moment when the first shock forms, are marked by the red crosses. The red line is a linear fitting to the marked strains, whose fitting expression has been shown in the figure.

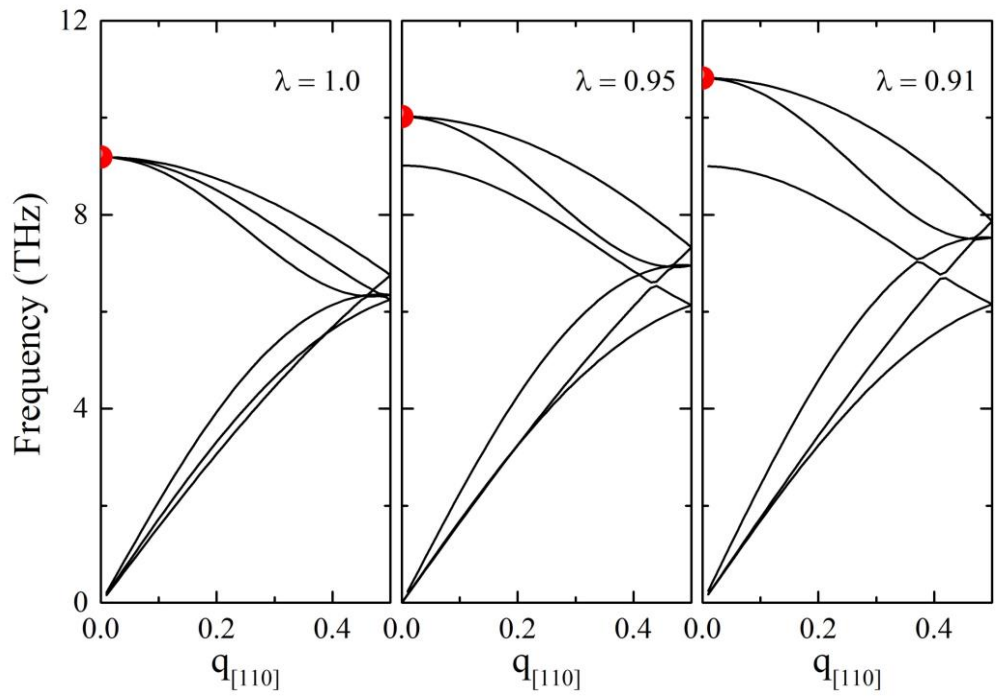

Fig. 12. Phonon dispersion curves of bcc iron under uniaxial compressions with different compression ratio, where the transverse optical phonons at the zone boundary are marked by red balls.

# Appendix A

In this part, new instability criteria will be developed for crystals under strain and strain gradient. In the main text, the developed instability criteria have been proved to be effective for judging instabilities of crystals under dynamic loadings. Now, imaging a small uniform strain gradient **κ** acting on a crystal, a linear strain field of **η(a)** is created in the crystal. Here, we use {**a**} and {**X**} to denote reference configuration and deformation configuration at the position of interest, respectively. Quantities with subscripts in upper case denote the one in {**X**}, while quantities with subscripts in lower case denote the one in {**a**}. Then free energy of a small volume centered at **X(a₀)** could be expressed as

$$F(\boldsymbol{\eta}, \boldsymbol{\kappa}, T) = \int \Big[ f(0,0,T) + \sigma_{ij}\eta_{ij} + \tau_{lmn}\kappa_{lmn} + \frac{1}{2}C_{ijmn}\eta_{ij}\eta_{mn} + W_{ijlmn}\eta_{ij}\kappa_{lmn} +$$

$$\frac{1}{2}T_{ijklmn}\kappa_{ijk}\kappa_{lmn}\Big]dV_{\mathbf{a}}, \tag{A.1}$$

where $f(0,0,T)$ is free energy density of unstrained lattice. Stress **σ**, higher order stress **τ** and the second order coefficients in (A.1) are defined by

$$\sigma_{ij} = \frac{1}{V_{\mathbf{a}}}\frac{\partial F}{\partial \eta_{ij}}\Big|_{\mathbf{a}}, \tag{A.2}$$

$$\tau_{lmn} = \frac{1}{V_{\mathbf{a}}}\frac{\partial F}{\partial \kappa_{lmn}}\Big|_{\mathbf{a}}, \tag{A.3}$$

$$C_{ijmn} = \frac{1}{V_{\mathbf{a}}}\frac{\partial^2 F}{\partial \eta_{ij}\partial \eta_{mn}}\Big|_{\mathbf{a}}, \tag{A.4}$$

$$W_{ijlmn} = \frac{1}{V_{\mathbf{a}}}\frac{\partial^2 F}{\partial \eta_{ij}\partial \kappa_{lmn}}\Big|_{\mathbf{a}}, \tag{A.5}$$

and

$$T_{ijklmn} = \frac{1}{V_{\mathbf{a}}}\frac{\partial^2 F}{\partial \kappa_{ijk}\partial \kappa_{lmn}}\Big|_{\mathbf{a}}. \tag{A.6}$$

According to the definition (1)-(3) in the main text, we have

$$X_{I,i} \approx \delta_{I,i} + \eta_{I,i}, \tag{A.7}$$

$$\det\left(\left[X_{I,i}\right]^{-1}\right) \approx 1 - \eta_{pp}, \tag{A.8}$$

where only the linear term of strain is retained. According to (A.1), the variation of the free energy in a small virtual strain gradient $\delta\kappa_{ijk}$ is, to the second order,

$$\delta F(\delta\boldsymbol{\eta}, \delta\boldsymbol{\kappa}, T) = \int \big[\sigma_{ij}\delta\eta_{ij} + \tau_{lmn}\delta\kappa_{lmn} + C_{ijmn}\delta\eta_{ij}\eta_{mn} + W_{ijlmn}\delta\eta_{ij}\kappa_{lmn} + W_{ijlmn}\eta_{ij}\delta\kappa_{lmn}$$

$$+ T_{ijklmn}\kappa_{ijk}\delta\kappa_{lmn}\big]dV_{\mathbf{a}}$$

$$= \int \big[\big(\sigma_{ij} + C_{ijmn}\eta_{mn} + W_{ijlmn}\kappa_{lmn}\big)X_{I,i}X_{J,j}\delta\eta_{IJ}$$

$$+ \big(\tau_{lmn} + W_{ijlmn}\eta_{ij} + T_{ijklmn}\kappa_{ijk}\big)X_{L,l}X_{M,m}X_{N,n}\delta\kappa_{LMN}\big]\det\left(\left[X_{I,i}\right]^{-1}\right)dV_{\mathbf{X}}$$

$$\tag{A.9}$$

where δ**η** is variation of strain at **X** due to the variation of the strain gradient. By substituting (A.7) and (A.8) into (A.9) and only retaining the second order terms of strain or strain gradient, we have

$$\delta F(\delta\boldsymbol{\eta},\delta\boldsymbol{\kappa},T) = \int [\sigma_{ij} + (C_{ijmn} + \sigma_{jn}\delta_{im} + \sigma_{in}\delta_{jm} - \sigma_{ij}\delta_{mn})\eta_{mn}$$
$$+ W_{ijlmn}\kappa_{lmn}]\delta_{I,i}\delta_{J,j}\delta\eta_{IJ}\mathrm{d}V_{\mathbf{X}}$$
$$+ \int [\tau_{lmn} + (W_{ijlmn} + \tau_{jmn}\delta_{il} + \tau_{ljn}\delta_{im} + \tau_{lmj}\delta_{in} - \tau_{lmn}\delta_{ij})\eta_{ij}$$
$$+ T_{ijklmn}\kappa_{ijk}]\delta_{L,l}\delta_{M,m}\delta_{N,n}\delta\kappa_{LMN}\mathrm{d}V_{\mathbf{X}}$$
$$= \int [\sigma_{IJ} + (C_{IJMN} + \sigma_{JN}\delta_{IM} + \sigma_{IN}\delta_{JM} - \sigma_{IJ}\delta_{MN})\eta_{MN} + W_{IJLMN}\kappa_{LMN}]\delta\eta_{IJ}\mathrm{d}V_{\mathbf{X}}$$
$$+ \int [\tau_{LMN} + (W_{IJLMN} + \tau_{JMN}\delta_{IL} + \tau_{LJN}\delta_{IM} + \tau_{LMJ}\delta_{IN} - \tau_{LMN}\delta_{IJ})\eta_{IJ}$$
$$+ T_{IJKLMN}\kappa_{IJK}]\delta\kappa_{LMN}\mathrm{d}V_{\mathbf{X}}$$
$$= \int [\sigma_{IJ} + (C_{IJMN} + \sigma_{JN}\delta_{IM} + \sigma_{IN}\delta_{JM} - \sigma_{IJ}\delta_{MN})\eta_{MN} + W_{IJLMN}\kappa_{LMN}]\delta\eta_{IJ}\mathrm{d}V_{\mathbf{X}}$$
$$+ \int [\tau_{LMN} + (W_{IJLMN} + \tau_{JMN}\delta_{IL} + \tau_{LJN}\delta_{IM} + \tau_{LMJ}\delta_{IN} - \tau_{LMN}\delta_{IJ})\eta_{IJ}$$
$$+ T_{IJKLMN}\kappa_{IJK}]\delta\kappa_{LMN}\mathrm{d}V_{\mathbf{X}}$$
$$= \int [\sigma_{IJ} + B_{IJMN}\eta_{MN} + W_{IJLMN}\kappa_{LMN}]\delta\eta_{IJ}\mathrm{d}V_{\mathbf{X}}$$
$$+ \int [\tau_{LMN} + \Lambda_{IJLMN}\eta_{IJ} + T_{IJKLMN}\kappa_{IJK}]\delta\kappa_{LMN}\mathrm{d}V_{\mathbf{X}}$$
$$= \int [\sigma_{IJ} + B_{IJMN}\eta_{MN} + W_{IJLMN}\kappa_{LMN}]\delta\eta_{IJ}\mathrm{d}V_{\mathbf{X}}$$
$$+ \int [\tau_{LMN} + \Lambda_{IJLMN}\eta_{IJ} + T_{IJKLMN}\kappa_{IJK}]\delta\eta_{MN}n_L\mathrm{d}S$$
$$- \int [\tau_{LMN} + \Lambda_{IJLMN}\eta_{IJ} + T_{IJKLMN}\kappa_{IJK}]_{,L}\delta\eta_{MN}\mathrm{d}V_{\mathbf{X}}$$
$$= \int \left[\sigma_{IJ} + B_{IJMN}\eta_{MN} + W_{IJLMN}\kappa_{LMN} - (\tau_{KIJ} + \Lambda_{MNKIJ}\eta_{MN} + T_{LMNKIJ}\kappa_{LMN})_{,K}\right]\delta\eta_{IJ}\mathrm{d}V_{\mathbf{X}}$$
$$+ \int [\tau_{LMN} + \Lambda_{IJLMN}\eta_{IJ} + T_{IJKLMN}\kappa_{IJK}]\delta\eta_{MN}n_L\mathrm{d}S$$
$$= \int [\sigma_{IJ} + \tau_{KIJ,K} + (B_{IJMN} - \Lambda_{MNKIJ,K})\eta_{MN} + (W_{IJLMN} - \Lambda_{MNLIJ} - T_{LMNKIJ,K})\kappa_{LMN}]\delta\eta_{IJ}\mathrm{d}V_{\mathbf{X}}$$
$$+ \int [\tau_{LMN} + \Lambda_{IJLMN}\eta_{IJ} + T_{IJKLMN}\kappa_{IJK}]\delta\eta_{MN}n_L\mathrm{d}S$$

(A.10)

where **n** is outward unit vector normal to surface (S) of $V_{\mathbf{X}}$ and

$$B_{IJKL} = C_{IJKL} + \sigma_{JL}\delta_{IK} + \sigma_{IL}\delta_{JK} - \sigma_{IJ}\delta_{KL},$$ (A.11)
$$\Lambda_{IJLMN} = W_{IJLMN} + \tau_{JMN}\delta_{IL} + \tau_{LJN}\delta_{IM} + \tau_{LMJ}\delta_{IN} - \tau_{LMN}\delta_{IJ}.$$ (A.12)

In the above derivations, divergence theorem and $\kappa_{IJK,L} = 0$ are used. In the last equation of (A.10), the second integration represents work done by $\tau_{LMN}$ due to the variation of strain gradients. Assuming the lattice is infinite, variation of free energy density, due to the variation of strain gradients, is

$$\delta f_{V_{\mathbf{X}}} = (\sigma_{IJ} + \tau_{KIJ,K})\delta\eta_{IJ} + (B_{IJMN} - \Lambda_{MNKIJ,K})\eta_{MN}\delta\eta_{IJ} + (W_{IJLMN} - \Lambda_{MNLIJ} - T_{LMNKIJ,K})\kappa_{LMN}\delta\eta_{IJ}.$$ (A.13)

Under equilibrium states, work done by external stresses is

$$\delta w = (\sigma_{IJ} + \tau_{KIJ,K})\delta\eta_{IJ}.$$ (A.14)

To make the crystal stable at **X**, the difference between the increment of free energy density and the work done by external stresses must be positive, that is

$$\delta f_{V_X} - \delta w = (B_{IJMN} - \Lambda_{MNKIJ,K})\eta_{MN}\delta\eta_{IJ} + (W_{IJLMN} - \Lambda_{MNLIJ} - T_{LMNKIJ,K})\kappa_{LMN}\delta\eta_{IJ} > 0. \tag{A.15}$$

Considering equation (A.12), condition (A.15) could reduce to

$$\begin{aligned}\delta f_{V_X} - \delta w &= (B_{IJMN} - \Lambda_{MNKIJ,K})\eta_{MN}\delta\eta_{IJ} \\ &\quad - (\tau_{JMN}\delta_{IL} + \tau_{LJN}\delta_{IM} + \tau_{LMJ}\delta_{IN} - \tau_{LMN}\delta_{IJ} + T_{LMNKIJ,K})\kappa_{LMN}\delta\eta_{IJ} \\ &= \tilde{B}_{IJMN}\eta_{MN}\delta\eta_{IJ} - \tilde{T}_{LMNIJ}\kappa_{LMN}\delta\eta_{IJ} > 0,\end{aligned} \tag{A.16}$$

by making use of $\sum_{IJMN} W_{IJLMN}\kappa_{LMN}\delta\eta_{IJ} = \sum_{IJMN} W_{MNLIJ}\kappa_{LMN}\delta\eta_{IJ}$. In above condition, we have defined

$$\tilde{B}_{IJMN} = B_{IJMN} - \Lambda_{MNKIJ,K}, \tag{A.17}$$

and

$$\tilde{T}_{LMNIJ} = T_{LMNKIJ,K} + \tau_{JMN}\delta_{IL} + \tau_{LJN}\delta_{IM} + \tau_{LMJ}\delta_{IN} - \tau_{LMN}\delta_{IJ}. \tag{A.18}$$

Because I and J (M and N) are exchangeable, (A.11), (A.12) and (A.18) could be written in a more symmetrical form as below:

$$B_{IJKL} = C_{IJKL} + \frac{1}{2}(\sigma_{JL}\delta_{IK} + \sigma_{IL}\delta_{JK} + \sigma_{JK}\delta_{IL} + \sigma_{IK}\delta_{JL} - 2\sigma_{IJ}\delta_{KL}), \tag{A.11'}$$

$$\Lambda_{IJLMN} = W_{IJLMN} + \frac{1}{2}(\tau_{JMN}\delta_{IL} + \tau_{LJN}\delta_{IM} + \tau_{LMJ}\delta_{IN} + \tau_{IMN}\delta_{JL} + \tau_{LIN}\delta_{JM} + \tau_{LMI}\delta_{JN} - 2\tau_{LMN}\delta_{IJ}). \tag{A.12'}$$

$$\tilde{T}_{LMNIJ} = T_{LMNKIJ,K} + \frac{1}{2}(\tau_{JMN}\delta_{IL} + \tau_{LJN}\delta_{IM} + \tau_{LMJ}\delta_{IN} + \tau_{IMN}\delta_{JL} + \tau_{LIN}\delta_{JM} + \tau_{LMI}\delta_{JN} - 2\tau_{LMN}\delta_{IJ}). \tag{A.18'}$$

Before discussing the stability condition at the presence of both strain and strain gradient, we revisit the stability condition without the strain gradient. In this case, $\tau_{JMN}$ and $W_{IJLMN}$ are zero when no strain gradient is presence (This would be discussed further in Appendix B). By substituting $\kappa_{LMN} = 0$ into (A.16), we get

$$\tilde{B}_{IJMN}\eta_{MN}\delta\eta_{IJ} = (B_{IJMN} - \Lambda_{MNKIJ,K})\eta_{MN}\delta\eta_{IJ} = B_{IJMN}\eta_{MN}\delta\eta_{IJ} > 0. \tag{A.19}$$

This is the stability condition at finite strain (Wang et al., 1993). When initial strain gradient exists, the stability condition (A.16) could be further expressed as

$$\tilde{B}_{RS}\tilde{\eta}_S\delta\tilde{\eta}_R - \tilde{T}_{LRS}\kappa_{LR}\delta\tilde{\eta}_S = \tilde{B}_{RS}\tilde{\eta}_S\delta\tilde{\eta}_R - \tilde{T}^L_{RS}\tilde{\kappa}_{LR}\delta\tilde{\eta}_S > 0, \tag{A.20}$$

where $R$ and $S$ (= 1, 2, 3, 4, 5, 6) obey Voigt convention, and $\tilde{\boldsymbol{\eta}}$ and $\tilde{\boldsymbol{\kappa}}$ relate to $\eta$ and $\kappa$ by

$$\begin{cases} \tilde{\eta}_I = \eta_{II}, & (I = 1,2,3) \\ \tilde{\eta}_4 = 2\eta_{23}, \tilde{\eta}_5 = 2\eta_{13}, \tilde{\eta}_6 = 2\eta_{12}, \end{cases} \tag{A.21}$$

and

$$\begin{cases} \tilde{\kappa}_{LI} = \kappa_{LII}, & (I = 1,2,3) \\ \tilde{\kappa}_{L4} = 2\kappa_{L23}, \tilde{\kappa}_{L5} = 2\kappa_{L13}, \tilde{\kappa}_{L6} = 2\kappa_{L12}. \end{cases} \quad (L = 1,2,3), \tag{A.22}$$

respectively. To obtain strain gradient stability criteria, $\tilde{\mathbf{T}}$ has been divided into three 6×6 *block matrixes* ($\tilde{\mathbf{T}}^L$). Therefore, for arbitrary $\tilde{\boldsymbol{\eta}}$ and $\tilde{\boldsymbol{\kappa}}$, absolute stability condition requires that $\tilde{\mathbf{B}}$ and $-\tilde{\mathbf{T}}^L$ ($L$=1, 2, 3) are positive definite. This condition could be used to judge stabilities of a crystal with finite initial strain and strain gradient. In practice, only *block matrix* along the direction of strain gradient needs to be considered when judging stabilities. Detailed physical meanings could be found in the main text.

# Appendix B

Because the new instability criteria in Appendix A are present in a continuum form, a continuum-consistence approach at atomic level should be designed in order to verify the correctness of the criteria. In this part, we will derive the higher order stress and the work conjugates of strain gradient tensor for simple lattices binding through embedded-atom-model (EAM) potential. If not specified, bold font letter stands for vector or tensor and otherwise, it represents the magnitude of the corresponding vector or merely a scalar. Lowercase Greek letters, such as $\alpha$, $\beta$, $\gamma$, $\mu$, $\nu$, $\lambda$ and $\rho$, are employed to distinguish the three Cartesian components of vectors or tensors, while lowercase English letters, such as $i$ and $j$, stand for atom indexes. Other conventions are the same as main text. Assuming that an infinite ideal lattice is deformed from reference configuration $\{X\}$ to current configuration $\{Y\}$ at position $X_0$ under a uniform strain gradient ($\kappa$), we will evaluate free energy density at the position after the deformation and further derive the quantities required by condition (A.20). Pairwise separation between atom $i$ and $j$ in configuration $\{X\}$ and $\{Y\}$ are described by $\mathbf{R}_{ij} = \mathbf{X}_i - \mathbf{X}_j$ and $\mathbf{r}_{ij} = \mathbf{Y}_i - \mathbf{Y}_j$, respectively. For brevity, we will write $r_{i0}$ ($R_{i0}$) as $r_i$ ($R_i$). Without loss of generality, we choose $\mathbf{X}_0$ and $\mathbf{Y}_0$ to be origin point. According to the above convention, we have $\mathbf{R}_{i0} = \mathbf{R}_i = \mathbf{X}_i$ ($\mathbf{r}_{i0} = \mathbf{r}_i = \mathbf{Y}_i$). Then, we proceed by expanding displacement of atom $i$ in terms of $\mathbf{X}_i$, that is

$$u_\alpha(\mathbf{X}_i) = u_\alpha(0) + u_{\alpha,\beta}(0)X_i^\beta + \frac{1}{2}u_{\alpha,\beta\gamma}(0)X_i^\beta X_i^\gamma, \tag{B.1}$$

where $u_\alpha(\mathbf{X}_i) = Y_i^\alpha - X_i^\alpha$ ($\alpha$, $\beta$, $\gamma$ = 1,2,3), $u_{\alpha,\beta}$ and $u_{\alpha,\beta\gamma}$ are the first and second gradient of the displacement. Third or higher order terms of $\mathbf{X}_i$ are not present in the expansion (B.1) because the strain gradient is uniform in space. According to the definition (1)-(3) in the main text, Lagrangian strain and strain gradient relate to the gradients of displacement by

$$\eta_{\alpha\beta} = \frac{1}{2}\left(u_{\alpha,\beta} + u_{\beta,\alpha} + u_{\mu,\alpha}u_{\mu,\beta}\right), \tag{B.2}$$

$$\kappa_{\gamma\alpha\beta} = \eta_{\alpha\beta,\gamma} = \frac{1}{2}\left(u_{\alpha,\beta\gamma} + u_{\beta,\alpha\gamma} + u_{\mu,\alpha\gamma}u_{\mu,\beta} + u_{\mu,\alpha}u_{\mu,\beta\gamma}\right). \tag{B.3}$$

The above equations are acquired without any approximations, especially (B.2) and (B.3), which will lead to different results compared with the ones under small linear strain approximation. The results will be shown later in this part. We need to be In the following part, we will derive an analytic expression of free energy density as a function of $\boldsymbol{\eta}$ and $\boldsymbol{\kappa}$. Firstly, considering that an infinite lattice is deformed by $\boldsymbol{\kappa}$, average energy density of a volume ($V$) centered at $\mathbf{X}_0$ could expressed by

$$\bar{e}_V(\boldsymbol{\eta}(\mathbf{X}_0), \boldsymbol{\kappa}) = \frac{1}{V}\sum_i^N U(\boldsymbol{\eta}(\mathbf{X}_i), \boldsymbol{\kappa}), \tag{B.4}$$

where $U(\boldsymbol{\eta}(\mathbf{X}_0), \boldsymbol{\kappa})$ represents the energy of atom $i$, and $N$ is atom number within the volume. It should be noted that $\bar{e}_V$ depends on V because of the existence of uniform strain gradient. Thus, it is necessary to define a characteristic size (or volume) before putting the stability condition (A.20) into practice at lattice level. From a microscopic point of view, instabilities of a lattice are usually triggered by fluctuations of atoms and would most probably first take place at position where the fluctuation is the largest. Then instabilities will develop rapidly from the position over the entire lattice and eventually lead to a phase transition or plasticity, for example melting (Yip et al., 2001), solid-solid phase transition (Wang et al., 2015) and yield (Krenn et al., 2001) of metals.

With this physical picture, we could define the characteristic volume to be average volume per atom in configuration {**X**} based on ideas that the instability of a crystal is first triggered by instabilities of an atom plane whose normal direction is parallel to the direction of strain gradient (See Fig. b1 for more details). Thereby, an average energy density over the characteristic volume centered at $\mathbf{X}_0$ is used to study the lattice instabilities, that is

$$\bar{e}_{\Omega_X}(\boldsymbol{\eta}, \boldsymbol{\kappa}) = \frac{1}{\Omega_X} U(\boldsymbol{\eta}, \boldsymbol{\kappa}) = \frac{1}{\Omega_X}[U_0 + \Delta U(\boldsymbol{\eta}, \boldsymbol{\kappa})], \tag{B.5}$$

where $\Omega_X$ is the atom volume at $\{\mathbf{X}_0\}$, $U_0$ is binding energy of the ideal lattice. The energy increment after introducing uniform strain gradient is

$$\Delta U(\boldsymbol{\eta}, \boldsymbol{\kappa}) = U(\boldsymbol{\eta}, \boldsymbol{\kappa}) - U_0, \tag{B.6}$$

where

$$U(\boldsymbol{\eta}, \boldsymbol{\kappa}) = \frac{1}{2}\sum_{i\neq 0}\phi(Y_i) + F(\rho_\mathbf{Y}) + M(P_\mathbf{Y}), \tag{B.7}$$

$$\rho_\mathbf{Y} = \sum_{i\neq 0} f(Y_i), \quad P_\mathbf{Y} = \sum_{i\neq 0} g(Y_i). \tag{B.8}$$

The detailed function form of pairwise interaction $\phi(r)$, embedding energy $F(\rho)$ and energy modified term $M(P)$ could be found in this reference (Wang et al., 2014). The summation runs over all neighbors of central atom located at $\mathbf{X}_0$ (or $\mathbf{Y}_0$). To express $\Delta U$ as a function of **u** and its derivatives, we assume that neighbors of the central atom do not change during the deformation. For presentation convenience, the assumption is referred to as unchanged-neighbor assumption (UNA) in this work. In atomic simulations, the neighbors of an atom are usually defined by a cutoff distance which is a parameter of EAM potentials. An atom is the neighbor of the central atom if the pairwise separation between them is smaller than (or equal to) the cutoff distance. Thus, the UNA means that the maximum displacement along the direction of strain gradient should not exceed

$$d_{max} = \text{MAX}\{R_{n+1}^e - R_c, R_c - R_n^e\}, \tag{B.9}$$

where $R_c$ is the cutoff distance and $R_{n+1}^e$ ($R_n^e$) is the $(n+1)$-th ($n$-th) neighbor separation of atoms at unstrained state satisfying $R_n^e < R_c < R_{n+1}^e$. For EAM potential of bcc iron (Wang et al., 2014), $n = 5$ and $R_c = R_5^e + 0.5(R_6^e - R_5^e)$. Then, we have $d_{max} = 0.5(R_6^e - R_5^e) = (1 - \sqrt{3}/2)a_0$, where $a_0$ is lattice constant of bcc iron, i.e., 2.8606 Å. The maximum displacement under uniform strain gradient (using $\kappa_m$ to denote the maximum component of **κ**) is about $u_{max} = \kappa_m(R_5^e)^2$. Thus, condition satisfied by the applied

$$\kappa_m < d_{max}/(R_5^e)^2 = (2 - \sqrt{3})/(6a_0) \approx 0.016 \text{ Å}^{-1}. \tag{B.10}$$

This condition is stricter than the facts. For certain direction of the strain gradient, allowed $d_{max}$ is larger than the one given in (B.9) because atoms do not necessary arrange along the direction. For example, if the direction of strain gradient is along $[001]_{bcc}$ direction, then $d_{max} = a_0/2$ and $u_{max} = \kappa(3a_0/2)^2$, which lead to a condition of $\kappa_m < 0.078 \text{ Å}^{-1}$. To reach the maximum strain gradient, we need to impact bcc iron with a strain rate of about $10^{12}$ s$^{-1}$! This condition is sufficient to satisfy most of cases encountered in engineering applications where the strain rates are usually not exceed $10^8$ s$^{-1}$.

With the UNA and equation (B.7)-(B.8), $\Delta U$ could be expanded as a function of **u** and its derivatives, that is

$$\Delta U(\boldsymbol{\eta}, \boldsymbol{\kappa}) = \frac{1}{2}\sum_{i\neq 0}[\phi(Y_i) - \phi(X_i)] + F(\rho_0(\{\mathbf{Y}_m\})) - F(\rho_0(\{\mathbf{X}_m\}))$$

$$+ M(P_0(\{\mathbf{Y}_m\})) - M(P_0(\{\mathbf{X}_m\}))$$

$$= \frac{1}{2}\sum_{i\neq 0} \Delta\phi(\mathbf{Y}_i) + \Delta F(\{\mathbf{Y}_m\}) + \Delta M(\{\mathbf{Y}_m\}), \tag{B.11}$$

where

$$\Delta\phi(\mathbf{Y}_i) = \phi(\mathbf{Y}_i) - \phi(\mathbf{X}_i)$$

$$= \phi'(X_i)\frac{X_i^\alpha}{X_i}u_i^\alpha + \frac{1}{2}\left[\phi''(X_i)\frac{X_i^\alpha X_i^\beta}{X_i^2} + \phi'(X_i)\left(\frac{1}{X_i}\delta_{\alpha\beta} - \frac{X_i^\alpha X_i^\beta}{X_i^3}\right)\right]u_i^\alpha u_i^\beta + \cdots, \tag{B.12}$$

$$\Delta F(\{\mathbf{Y}_m\}) = \frac{\partial F}{\partial \rho_0}\Delta\rho_0 + \frac{1}{2}\frac{\partial^2 F}{\partial \rho_0^2}(\Delta\rho_0)^2 + \cdots \tag{B.13}$$

$$\Delta M(\{\mathbf{Y}_m\}) = \frac{\partial P}{\partial g_0}\Delta g_0 + \frac{1}{2}\frac{\partial^2 P}{\partial g_0^2}(\Delta g_0)^2 + \cdots \tag{B.14}$$

And increments of $\rho_0$ and $g_0$ due to disturbance of the strain gradient are expanded to the second order of displacements, that is

$$\Delta\rho_0 = \sum_{i\neq 0}[f(\mathbf{Y}_i) - f(\mathbf{X}_i)] = \sum_{i\neq 0}\left[\frac{\partial f}{\partial X_i^\alpha}u_i^\alpha + \frac{1}{2}\frac{\partial^2 f}{\partial X_i^\alpha \partial X_i^\beta}u_i^\alpha u_i^\beta\right]$$

$$= \sum_{i\neq 0}\left[f'(X_i)\frac{X_i^\alpha}{X_i}u_i^\alpha + \frac{1}{2}\left(f''(X_i)\frac{X_i^\alpha X_i^\beta}{X_i^2} + f'(X_i)\left(\frac{1}{X_i}\delta_{\alpha\beta} - \frac{X_i^\alpha X_i^\beta}{X_i^3}\right)\right)u_i^\alpha u_i^\beta\right], \tag{B.15}$$

$$\Delta g_0 = \sum_{i\neq 0}[g(\mathbf{Y}_i) - g(\mathbf{X}_i)] = \sum_{i\neq 0}\left[\frac{\partial g}{\partial X_i^\alpha}u_i^\alpha + \frac{1}{2}\frac{\partial^2 g}{\partial X_i^\alpha \partial X_i^\beta}u_i^\alpha u_i^\beta\right]$$

$$= \sum_{i\neq 0}\left[g'(X_i)\frac{X_i^\alpha}{X_i}u_i^\alpha + \frac{1}{2}\left(g''(X_i)\frac{X_i^\alpha X_i^\beta}{X_i^2} + g'(X_i)\left(\frac{1}{X_i}\delta_{\alpha\beta} - \frac{X_i^\alpha X_i^\beta}{X_i^3}\right)\right)u_i^\alpha u_i^\beta\right]. \tag{B.16}$$

Substituting equation (B.15) and (B.16) into (B.13) and (B.14), respectively, we get

$$\Delta F(\{\mathbf{Y}_m\}) =$$

$$F'(\rho_0)\sum_{i\neq 0}\left[f'(X_i)\frac{X_i^\alpha}{X_i}u_i^\alpha + \frac{1}{2}\left(f''(X_i)\frac{X_i^\alpha X_i^\beta}{X_i^2} + f'(X_i)\left(\frac{1}{X_i}\delta_{\alpha\beta} - \frac{X_i^\alpha X_i^\beta}{X_i^3}\right)\right)u_i^\alpha u_i^\beta\right] +$$

$$\frac{1}{2}F''(\rho_0)\sum_{i\neq 0}\sum_{j\neq 0}f'(X_i)f'(X_j)\frac{X_i^\alpha X_j^\beta}{X_i X_j}u_i^\alpha u_j^\beta, \tag{B.17}$$

$$\Delta M(\{\mathbf{Y}_m\}) = \frac{1}{N_0}\sum_i\left[M(P_i(\{\mathbf{Y}\})) - M(P_i(\{\mathbf{X}\}))\right]$$

$$= M'(P_0)\sum_{i\neq 0}\left[g'(X_i)\frac{X_i^\alpha}{X_i}u_i^\alpha + \frac{1}{2}\left(g''(X_i)\frac{X_i^\alpha X_i^\beta}{X_i^2} + g'(X_i)\left(\frac{1}{X_i}\delta_{\alpha\beta} - \frac{X_i^\alpha X_i^\beta}{X_i^3}\right)\right)u_i^\alpha u_i^\beta\right] +$$

$$\frac{1}{2}M''(P_0)\sum_{i\neq 0}\sum_{j\neq 0}g'(X_i)g'(X_j)\frac{X_i^\alpha X_j^\beta}{X_i X_j}u_i^\alpha u_j^\beta. \tag{B.18}$$

Using equation (B.12), (B.17) and (B.18), equation (B.11) could be rewritten as

$$\Delta U(\boldsymbol{\eta}, \boldsymbol{\kappa}) = \frac{1}{2}\sum_{i\neq 0}\left\{\phi'(X_i)\frac{X_i^\alpha}{X_i}u_i^\alpha + \frac{1}{2}\left[\phi''(X_i)\frac{X_i^\alpha X_i^\beta}{X_i^2} + \phi'(X_i)\left(\frac{1}{X_i}\delta_{\alpha\beta} - \frac{X_i^\alpha X_i^\beta}{X_i^3}\right)\right]u_i^\alpha u_i^\beta\right\} +$$

$$\sum_{i\neq 0}F'(\rho_0)f'(X_i)\frac{X_i^\alpha}{X_i}u_i^\alpha + \frac{1}{2}\sum_{i\neq 0}\sum_{j\neq 0}F''(\rho_0)f'(X_i)f'(X_j)\frac{X_i^\alpha X_j^\beta}{X_i X_j}u_i^\alpha u_j^\beta +$$

$$\frac{1}{2}\sum_{i\neq 0}F'(\rho_0)\left(f''(X_i)\frac{X_i^\alpha X_i^\beta}{X_i^2} + f'(X_i)\left(\frac{1}{X_i}\delta_{\alpha\beta} - \frac{X_i^\alpha X_i^\beta}{X_i^3}\right)\right)u_i^\alpha u_i^\beta + \sum_{i\neq 0}M'(P_0)g'(X_i)\frac{X_i^\alpha}{X_i}u_i^\alpha +$$

$\frac{1}{2}\sum_{i\neq 0}\sum_{j\neq 0}M''(P_0)g'(X_i)g'(X_j)\frac{X_i^\alpha X_j^\beta}{X_i X_j}u_i^\alpha u_j^\beta + \frac{1}{2}\sum_{i\neq 0}M'(P_0)\left(g''(X_i)\frac{X_i^\alpha X_i^\beta}{X_i^2} + g'(X_i)\left(\frac{1}{X_i}\delta_{\alpha\beta} - \frac{X_i^\alpha X_i^\beta}{X_i^3}\right)\right)u_i^\alpha u_i^\beta$

$= \sum_{i\neq 0}[\phi'(X_i)/2 + F'f'(X_i) + M'g'(X_i)]\frac{X_i^\alpha}{X_i}u_i^\alpha$

$+ \frac{1}{2}\sum_{i\neq 0}\left\{\frac{1}{2}\left[\phi''(X_i)\frac{X_i^\alpha X_i^\beta}{X_i^2} + \phi'(X_i)\left(\frac{1}{X_i}\delta_{\alpha\beta} - \frac{X_i^\alpha X_i^\beta}{X_i^3}\right)\right] + F'\left[f''(X_i)\frac{X_i^\alpha X_i^\beta}{X_i^2} + f'(X_i)\left(\frac{1}{X_i}\delta_{\alpha\beta} - \frac{X_i^\alpha X_i^\beta}{X_i^3}\right)\right] + M'\left[g''(X_i)\frac{X_i^\alpha X_i^\beta}{X_i^2} + g'(X_i)\left(\frac{1}{X_i}\delta_{\alpha\beta} - \frac{X_i^\alpha X_i^\beta}{X_i^3}\right)\right]\right\}u_i^\alpha u_i^\beta + \frac{1}{2}\sum_{i\neq 0}\sum_{j\neq 0}[F''f'(X_i)f'(X_j) + M''g'(X_i)g'(X_j)]\frac{X_j^\beta}{X_j}\frac{X_i^\alpha}{X_i}u_i^\alpha u_j^\beta$

$\equiv \sum_{i\neq 0}P_i^\alpha u_i^\alpha + \frac{1}{2}\sum_{i\neq 0}V_i^{\alpha\beta}u_i^\alpha u_i^\beta + \frac{1}{2}\sum_{i\neq 0}\sum_{j\neq 0}H_{ij}^{\alpha\beta}u_i^\alpha u_j^\beta,$ (B.19)

where

$P_i^\alpha = [\phi'(X_i)/2 + F'f'(X_i) + M'g'(X_i)]\frac{X_i^\alpha}{X_i},$ (B.20)

$V_i^{\alpha\beta} = \frac{1}{2}\left[\phi''(X_i)\frac{X_i^\alpha X_i^\beta}{X_i^2} + \phi'(X_i)\left(\frac{1}{X_i}\delta_{\alpha\beta} - \frac{X_i^\alpha X_i^\beta}{X_i^3}\right)\right] + F'\left[f''(X_i)\frac{X_i^\alpha X_i^\beta}{X_i^2} + f'(X_i)\left(\frac{1}{X_i}\delta_{\alpha\beta} - \frac{X_i^\alpha X_i^\beta}{X_i^3}\right)\right] + M'\left[g''(X_i)\frac{X_i^\alpha X_i^\beta}{X_i^2} + g'(X_i)\left(\frac{1}{X_i}\delta_{\alpha\beta} - \frac{X_i^\alpha X_i^\beta}{X_i^3}\right)\right],$ (B.21)

$H_{ij}^{\alpha\beta} = [F''f'(X_i)f'(X_j) + M''g'(X_i)g'(X_j)]\frac{X_i^\alpha}{X_i}\frac{X_j^\beta}{X_j}.$ (B.22)

Substituting equation (B.1) into (B.19) and making use of $u_\alpha(0) = 0$ (See Fig. b1), we have

$\Delta U(\boldsymbol{\eta},\boldsymbol{\kappa}) = \sum_{i\neq 0}P_i^\alpha\left(u_{\alpha,\mu}X_i^\mu + \frac{1}{2}u_{\alpha,\mu\nu}X_i^\mu X_i^\nu\right)$

$+ \frac{1}{2}\sum_{i\neq 0}V_i^{\alpha\beta}\left(u_{\alpha,\mu}X_i^\mu + \frac{1}{2}u_{\alpha,\mu\nu}X_i^\mu X_i^\nu\right)\left(u_{\beta,\lambda}X_i^\lambda + \frac{1}{2}u_{\beta,\lambda\rho}X_i^\lambda X_i^\rho\right)$

$+ \frac{1}{2}\sum_i\sum_j H_{ij}^{\alpha\beta}\left(u_{\alpha,\mu}X_i^\mu + \frac{1}{2}u_{\alpha,\mu\nu}X_i^\mu X_i^\nu\right)\left(u_{\beta,\lambda}X_j^\lambda + \frac{1}{2}u_{\beta,\lambda\rho}X_j^\lambda X_j^\rho\right)$

$= \left(\sum_{i\neq 0}P_i^\alpha X_i^\mu\right)u_{\alpha,\mu} + \frac{1}{2}\left(\sum_{i\neq 0}P_i^\alpha X_i^\mu X_i^\nu\right)u_{\alpha,\mu\nu}$

$+ \frac{1}{2}\left(\sum_{i\neq 0}V_i^{\alpha\beta}X_i^\mu X_i^\lambda + \sum_i\sum_j H_{ij}^{\alpha\beta}X_i^\mu X_j^\lambda\right)u_{\alpha,\mu}u_{\beta,\lambda}$

$+ \frac{1}{4}\left(\sum_{i\neq 0}V_i^{\alpha\beta}X_i^\mu X_i^\lambda X_i^\rho + \sum_i\sum_j H_{ij}^{\alpha\beta}X_i^\mu X_j^\lambda X_j^\rho\right)u_{\alpha,\mu}u_{\beta,\lambda\rho}$

$+ \frac{1}{4}\left(\sum_{i\neq 0}V_i^{\alpha\beta}X_i^\mu X_i^\nu X_i^\lambda + \sum_i\sum_j H_{ij}^{\alpha\beta}X_i^\mu X_i^\nu X_j^\lambda\right)u_{\alpha,\mu\nu}u_{\beta,\lambda}$

$+ \frac{1}{8}\left(\sum_{i\neq 0}V_i^{\alpha\beta}X_i^\mu X_i^\nu X_i^\lambda X_i^\rho + \sum_i\sum_j H_{ij}^{\alpha\beta}X_i^\mu X_i^\nu X_j^\lambda X_j^\rho\right)u_{\alpha,\mu\nu}u_{\beta,\lambda\rho}$

$= \tilde{P}_{\alpha\mu}u_{\alpha,\mu} + \tilde{Q}_{\alpha\mu\nu}u_{\alpha,\mu\nu} + \frac{1}{2}\tilde{C}_{\alpha\mu\beta\lambda}u_{\alpha,\mu}u_{\beta,\lambda} + \tilde{V}_{\alpha\mu\beta\lambda\rho}u_{\alpha,\mu}u_{\beta,\lambda\rho}$

$$+\frac{1}{2}\widetilde{H}_{\alpha\mu\nu\beta\lambda\rho}u_{\alpha,\mu\nu}u_{\beta,\lambda\rho}, \tag{B.23}$$

where

$$\widetilde{P}_{\alpha\mu} = \sum_{i\neq 0} P_i^{\alpha} X_i^{\mu}, \tag{B.24}$$

$$\widetilde{Q}_{\alpha\mu\nu} = \frac{1}{2}\sum_{i\neq 0} P_i^{\alpha} X_i^{\mu} X_i^{\nu}, \tag{B.25}$$

$$\widetilde{C}_{\alpha\mu\beta\lambda} = \sum_{i\neq 0} V_i^{\alpha\beta} X_i^{\mu} X_i^{\lambda} + \sum_i \sum_j H_{ij}^{\alpha\beta} X_i^{\mu} X_j^{\lambda}, \tag{B.26}$$

$$\widetilde{V}_{\alpha\mu\beta\lambda\rho} = \frac{1}{4}\left(2\sum_{i\neq 0} V_i^{\alpha\beta} X_i^{\mu} X_i^{\lambda} X_i^{\rho} + \sum_i \sum_j H_{ij}^{\alpha\beta} X_i^{\mu} X_j^{\lambda} X_j^{\rho} + \sum_i \sum_j H_{ij}^{\alpha\beta} X_i^{\lambda} X_j^{\rho} X_j^{\mu}\right), \tag{B.27}$$

$$\widetilde{H}_{\alpha\mu\nu\beta\lambda\rho} = \frac{1}{4}\left(\sum_{i\neq 0} V_i^{\alpha\beta} X_i^{\mu} X_i^{\nu} X_i^{\lambda} X_i^{\rho} + \sum_i \sum_j H_{ij}^{\alpha\beta} X_i^{\mu} X_i^{\nu} X_j^{\lambda} X_j^{\rho}\right). \tag{B.28}$$

Using (B.2) and (B.3) to convert the displacement gradients of equation (B.23) into stain and strain gradients and omitting terms higher than the second order, we get

$$\Delta U(\boldsymbol{\eta}, \boldsymbol{\kappa}) = \Omega_{\mathbf{X}}\left(\sigma_{\alpha\mu}\eta_{\alpha\mu} + \frac{1}{2}C_{\alpha\mu\beta\lambda}\eta_{\alpha\mu}\eta_{\beta\lambda} + \tau_{\nu\alpha\mu}\kappa_{\nu\alpha\mu} + W_{\alpha\mu\rho\beta\lambda}\eta_{\alpha\mu}\kappa_{\rho\beta\lambda} + \frac{1}{2}T_{\nu\alpha\mu\rho\beta\lambda}\kappa_{\nu\alpha\mu}\kappa_{\rho\beta\lambda}\right), \tag{B.29}$$

where

$$\sigma_{\alpha\mu} = \widetilde{P}_{\alpha\mu}/\Omega_{\mathbf{X}}, \tag{B.30}$$

$$C_{\alpha\mu\beta\lambda} + C_{\alpha\mu\lambda\beta} + C_{\mu\alpha\beta\lambda} + C_{\mu\alpha\lambda\beta} = 4\left(\widetilde{C}_{\alpha\mu\beta\lambda} - \delta_{\alpha\beta}\widetilde{P}_{\mu\lambda}\right)/\Omega_{\mathbf{X}}, \tag{B.31}$$

$$\tau_{\nu\alpha\mu} = \widetilde{Q}_{\alpha\mu\nu}/\Omega_{\mathbf{X}}, \tag{B.32}$$

$$W_{\alpha\mu\rho\beta\lambda} + W_{\mu\alpha\rho\beta\lambda} + W_{\alpha\mu\rho\lambda\beta} + W_{\mu\alpha\rho\lambda\beta} = 4\left(\widetilde{V}_{\alpha\mu\beta\lambda\rho} - \widetilde{Q}_{\lambda\mu\rho}\delta_{\alpha\beta}\right)/\Omega_{\mathbf{X}}, \tag{B.33}$$

$$T_{\nu\alpha\mu\rho\beta\lambda} + T_{\nu\alpha\mu\rho\lambda\beta} + T_{\nu\mu\alpha\rho\beta\lambda} + T_{\nu\mu\alpha\rho\lambda\beta} = 4\widetilde{H}_{\alpha\mu\nu\beta\lambda\rho}/\Omega_{\mathbf{X}}. \tag{B.34}$$

Due to the symmetries of ($\alpha\leftrightarrow\mu$), ($\beta\leftrightarrow\lambda$) and ($\alpha\mu$)$\leftrightarrow$($\beta\lambda$), (B.31), (B.33) and (B.34) could be rewritten as

$$\Omega_{\mathbf{X}} C_{\alpha\mu\beta\lambda} = \widetilde{C}_{\alpha\mu\beta\lambda} - \delta_{\alpha\beta}\widetilde{P}_{\mu\lambda}, \tag{B.31'}$$

$$\Omega_{\mathbf{X}} W_{\alpha\mu\rho\beta\lambda} = \widetilde{V}_{\alpha\mu\beta\lambda\rho} - \widetilde{Q}_{\lambda\mu\rho}\delta_{\alpha\beta}, \tag{B.33'}$$

$$\Omega_{\mathbf{X}} T_{\nu\alpha\mu\rho\beta\lambda} = \widetilde{H}_{\alpha\mu\nu\beta\lambda\rho}. \tag{B.34'}$$

Notably, $C_{\alpha\mu\beta\lambda}$ and $W_{\alpha\mu\rho\beta\lambda}$ are not equal to the ones under small linear strain approximation, which are $\widetilde{C}_{\alpha\mu\beta\lambda}/\Omega_{\mathbf{X}}$ and $\widetilde{V}_{\alpha\mu\beta\lambda\rho}/\Omega_{\mathbf{X}}$, respectively. This is because the second order term in definition (B.2) and (B.3) are retained in our derivations. If $C_{\alpha\mu\beta\lambda}$ and $W_{\alpha\mu\rho\beta\lambda}$ is converted to the strained configuration {**Y**} by utilizing relation (A.7), we will find that they are indeed **B** and **Λ** defined by equation (A.11) and (A.12), respectively. Then, through substituting (B.23) into (B.5), the energy density at $\mathbf{X}_0$ is

$$e_{\Omega_X}(\boldsymbol{\eta}, \boldsymbol{\kappa}) = \frac{1}{\Omega_X}[U_0 + \Delta U(\boldsymbol{\eta}, \boldsymbol{\kappa})]$$

$$= e_0 + \sigma_{\alpha\mu}\eta_{\alpha\mu} + \frac{1}{2}C_{\alpha\mu\beta\lambda}\eta_{\alpha\mu}\eta_{\beta\lambda} + \tau_{\nu\alpha\mu}\kappa_{\nu\alpha\mu} + W_{\alpha\mu\rho\beta\lambda}\eta_{\alpha\mu}\kappa_{\rho\beta\lambda} + \frac{1}{2}T_{\nu\alpha\mu\rho\beta\lambda}\kappa_{\nu\alpha\mu}\kappa_{\rho\beta\lambda}, \tag{B.35}$$

where $e_0$ is energy density of the ideal lattice. Equation (B.33) could be viewed as a second order expansion of Helmholtz free energy at 0 K with respective to strain and strain gradient, where **σ**, **C**, **τ**, **W** and **T** are stress, elastic stiffness tensor, higher order stress, the first and the second order work conjugates of strain gradient. Therefore, we have obtained expressions of the higher order stresses and the second order coefficients of strain gradient tensor under the framework of embedded-atom method. Although the above derivation of equation (B.35) is under a zero strain configuration, the same procedures could be applied for initial lattice under uniform strain. In this case, all quantities in the right hand side of equation (B.35) are corresponding to the ones measured with respective to the uniform strained configuration.

Interestingly, to evaluate $\boldsymbol{\tau}$, $\mathbf{W}$ and $\mathbf{T}$ through equation (B.32)-(B.34), we only need to know the strain ($\boldsymbol{\eta}$) of a lattice without zero strain gradient. In other words, $\boldsymbol{\tau}$, $\mathbf{W}$ and $\mathbf{T}$ is only a function of $\boldsymbol{\eta}$ under the UNA. This is because that the concept of strain gradient could only exist when there are more than one primitive lattice cells in the characteristic volume, while our characteristic volume only contains one primitive lattice cell. This is a result of spatial discretization at lattice level. However, as shown in (B.35), the energy density over the characteristic volume could be influenced by the strain gradient because the potential cutoff distance is usually larger than a lattice constant (which is referred to as strain gradient effects in present work). It is need to point out that strain gradients, caused by usual collision or impacting (typical strain rate of $< 10^6$ s$^{-1}$), are not enough to affect local lattice within a volume characterized by the cutoff distance. With the increasing strain rate, the strain gradient effects will become obvious and cannot be neglected. Besides, according to equation (B.32) and (B.33) (as well as related expressions required by these two equations), $\boldsymbol{\tau}$ and $\mathbf{W}$ are zero if crystals are central-symmetric. Fortunately, all simple lattices, described by a Bravais lattice, are central-symmetric. In this case, equation (A.18) could be further written as

$$\tilde{T}_{LMNIJ} = T_{LMNKIJ,K} = \frac{\partial T_{LMNKIJ}}{\partial X_K} = \frac{\partial T_{LMNKIJ}}{\partial \eta_{PQ}} \kappa_{KPQ}, \tag{B.36}$$

or

$$\tilde{T}_{RS}^L = \frac{\partial T_{LRKS}}{\partial \tilde{\eta}_Z} \tilde{\kappa}_{KZ}, \tag{B.37}$$

where $L$, $M$, $N$, $I$, $J$, $K$ (= 1, 2, 3) are Cartesian indexes, $R$, $S$, $Z$ (= 1, 2, 3, 4, 5, 6) obey Voigt notation.

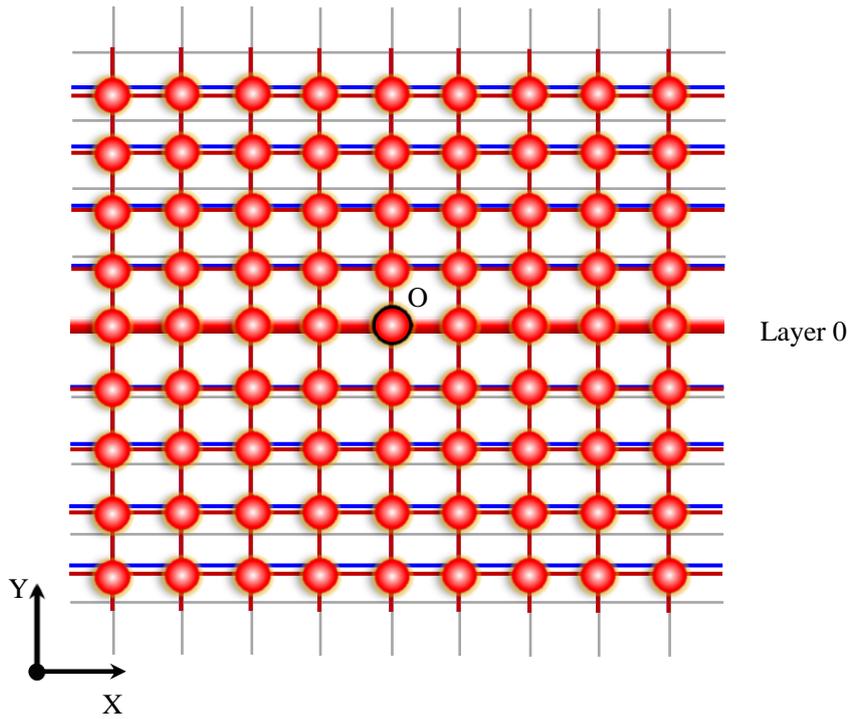

Fig. b1. A deformed 2D-lattice under uniform strain gradient, where the gray grid and the blue grid represent the ideal lattice and strained lattice under uniform uniaxial strain, respectively. The strain gradient is applied along Y direction. If local lattice element centered at "O" (as marked in the figure) is unstable, then all local lattice elements (forming an atom plane marked by bold red line), whose centers locate in layer 0, are unstable. If we choose "O" as our origin point, displacement of atom at the origin point is always zero during the deformations.